\documentclass[3p]{elsarticle} 
 \myfooter[C]{}
\renewcommand{\thispagestyle}[1]{}

\makeatother

\date{}
\def\texpsfig#1#2#3{\vbox{\kern #3\hbox{\includegraphics{#1}\kern #2}}\typeout{(#1)}}

\usepackage{url,hyperref}
\hypersetup{
    colorlinks=true,
    linkcolor=blue,
    filecolor=magenta,
    urlcolor=cyan,
    pdfpagemode=FullScreen,
    }
\usepackage{bbm}
\usepackage{eurosym}                           
\usepackage[latin1]{inputenc}                  
\usepackage{url}                               
\usepackage{longtable}                         
\usepackage{array}                             
\usepackage{graphicx,color}
\usepackage{amsthm}
\usepackage{amsbsy}

\usepackage{amssymb}
\usepackage{amsmath}
\usepackage{enumerate}
\usepackage{graphicx,color}
\usepackage{epsfig}
\usepackage[ruled,vlined]{algorithm2e}
\usepackage{multirow,bigdelim}
\usepackage[table]{xcolor}
\theoremstyle{plain}
\newtheorem{thm}{Theorem}[section]
\newtheorem{cor}{Corollary}[section]

\newtheorem{rem}[thm]{Remark}
\theoremstyle{remark}

\theoremstyle{plain}
\newtheorem{lem}[thm]{Lemma}

\newtheorem{prop}[thm]{Proposition}

\theoremstyle{definition}

\newcommand{\e}{{\rm e}}        
\def\R{\mathbb{ R}}             
\def\E{\mathbb{ E}}             
\def\Q{\mathbb{ Q}}             

\def\F{\mathcal{F}}             

\renewcommand{\d}{{\rm d}}      
\def\dW{{\rm d}W}               
\def\dt{{\rm d}t}
\def\ds{{\rm d}s}

\def\dz{{\rm d}z}

\def\T{{\rm T}}
\def\rHW{{\rm rHW}}
\def\HW{{\rm HW}}
\def\1{{\mathbbm{1}}}            

\theoremstyle{plain}

\newtheorem{theorem}{Theorem}[section]

\usepackage[margin=1cm]{caption}

\numberwithin{equation}{section}	     

\title{\raggedright Randomization of Short-Rate Models, Analytic Pricing and Flexibility in Controlling Implied Volatilities}

\begin{document}

\author[1,2]{\raggedright LECH A. GRZELAK\corref{cor1}}
\ead{L.A.Grzelak@uu.nl}
\cortext[cor1]{Corresponding author at Mathematical Institute, Utrecht University, Utrecht, the Netherlands.}
\address[1]{Mathematical Institute, Utrecht University, Utrecht, the Netherlands}
\address[2]{Financial Engineering, Rabobank, Utrecht, the Netherlands}
\begin{abstract}
We focus on extending existing short-rate models, enabling control of the generated implied volatility while preserving analyticity. We achieve this goal by applying the Randomized Affine Diffusion (RAnD) method~\cite{grzelakRAnD} to the class of short-rate processes under the Heath-Jarrow-Morton framework. Under arbitrage-free conditions, the model parameters can be exogenously stochastic, thus facilitating additional degrees of freedom that enhance the calibration procedure. We show that with the randomized short-rate models, the shapes of implied volatility can be controlled and significantly improve the quality of the model calibration, even for standard 1D variants. In particular, we illustrate that randomization applied to the Hull-White model leads to dynamics of the local volatility type, with the prices for standard volatility-sensitive derivatives explicitly available. The randomized Hull-White (rHW) model offers an almost perfect calibration fit to the swaption implied volatilities.
\noindent
\end{abstract}

\maketitle
{\let\thefootnote\relax\footnotetext{The views expressed in this paper are the author's personal views and do not necessarily reflect the views or policies of his current or past employers. The author has no competing interests.}}


\section{Introduction}
Choosing a pricing model, especially in risk management and for Valuation Adjustments (xVA), is a balancing act between accuracy and pricing speed. Often, the calculation speed is critical in determining the model of choice, especially when handling large portfolios. The modern approach for modelling interest rates is based on the Heath, Jarrow, and Morton (HJM)~\cite{HJM:1992} framework with an arbitrary term structure of volatility and covariance of {\it forward rates} across maturities. The framework is generic, and essentially any sensible term structure dynamics can be modelled by the HJM technique. Unfortunately, the number of models giving rise to analytical formulae for option pricing or even zero-coupon bonds is minimal. Therefore, models that can generate realistic implied volatilities while allowing for efficient derivative pricing are in high demand, especially when the portfolios need to be evaluated multiple times and for different market scenarios due to regulatory pressure.

A family of the HJM models that has gained particular acceptance from practitioners and academia is the family of affine models~\cite{Duffie:2000}. Over many years, various affine models have been studied, leading to significant contributions regarding efficient pricing, simulation and calibration (for the overview, see~\cite{BrigoMercurio:2007}). In particular, the classic affine short-rate models, like the Hull-White~\cite{HullWhite:1990} model, are popularised due to the closed formula for zero-coupon bonds and semi-analytic swaption pricing. On the downside, they lack sufficient flexibility to calibrate to market-implied volatilities. Additional model flexibility without compromising numerical efficiency is highly desired. A straightforward approach to incorporate and control the implied volatility smile and skew into the short-rate models is to define a stochastic or a local volatility process, for example, in~\cite{casassus2005unspanned,fong1991fixed,gatarek2017nonparametric}. Although such models give us a closed-form expression for the zero-coupon bond, pricing more advanced derivatives like options often involves advanced numerical techniques like Fourier inversion or numerical integration, which is undesired for the pricing of high-volume derivatives. A survey of the most popular stochastic volatility models in the interest rate world can be found in~\cite{andreasen2010stochastic}.

This article focuses on extending existing interest rate models with model parameters to be random via the so-called randomization method (RAnD)~\cite{grzelakRAnD}. Randomization by making the model parameters stochastic is a technique for more flexibility in the stochastic model-thus, improving the calibration quality while preserving the analytic properties of the models considered. The concept of randomization represents the uncertainty of potentially hidden states that are not sufficiently captured by deterministic parameters. Conventionally, under the affine framework, an extension by a stochastic parameter would require the model to meet linearity conditions - this does not have to be the case under RAnD. The method requires an affinity to hold, given a specific realization of the stochastic parameter. In other words, it builds an outer layer over the affine models and allows a stochastic parameter setting in that layer. The quadrature rule handles the numerical complexity associated with an infinite number of parameter values and reduces it to only a few {\it critical} parameter realizations. The selection of these points is based on the moments of the stochastic parameter.

The RAnD method is generic and can be applied to any pricing model. For the sake of simplicity, however, we focus on one of the most popular interest rate models, namely the Hull-White (HW) model~\cite{HullWhite:1990}. In particular, we will show that applying the RAnD method to the HW model results in a local-volatility type of dynamics. The local volatility dynamics may seem like a ``deal-breaker''; typically, the local volatility models, especially in interest rates, are notoriously tricky to operate, i.e., closed-form option pricing is hardly possible. However, although the randomized Hull-White (rHW) model is of the local volatility type, the option pricing is as complex as the standard, {\it unrandomized} model. Furthermore, the presented model offers fast and accurate calibration as the local volatility dynamics provide more flexibility in generating volatility smiles and skews for various market conditions.

Our method also addresses a challenging problem stated by Brigo and Mercurio in~\cite{brigo2002lognormal}, where the mixture of lognormals was studied. In their paper, it was shown that great flexibility to standard models like Black-Scholes was introduced by considering a convex combination of the associated probability density functions: $\omega_1f_{1}(x;\sigma_1)+\dots+\omega_Nf_{N}(x;\sigma_N)$ with $\omega_1+\dots+\omega_N=1$ and some $\sigma_i>0$, $i=1,\dots,N.$ However, the problem encountered was the large number of the model parameters that are difficult to interpret and relate to the corresponding implied volatilities. As stated in their work: {``\it the absence of bounds on the parameter $N$ implies that a virtually unlimited number of parameters can be introduced in the dynamics so as to be used for a better calibration to market data. (...), one has to find the correct tradeoff between model flexibility and number of parameters so as to avoid both poor calibration and over-parametrization issues''}. However, this problem is resolved with the RAnD method. The method transitions from a continuous randomizer $\vartheta$ to a unique sequence of weights $\omega_i$ and associated values $\sigma_i$. In this setting, the model calibration is performed by varying the parameters of the randomizer, $\vartheta$; therefore, the number of free parameters is manageable. Moreover, as presented in this work, the varying parameters of $\vartheta$ enable a transparent control of the implied volatilities.

This article comprises five sections. First, the details of the HJM framework and pricing equations using the RAnD method are provided in Section~\ref{sec:HJM}, and a discussion on the arbitrage-free conditions is covered in Section~\ref{sec:arbitrageHJM}. In Section~\ref{sec:randomizedHW}, we consider the rHW model and derive the dynamics of the randomized stochastic processes. Section~\ref{sec:dynamics_rHW} discusses different variants of the rHW model parameter choices, namely the randomization using univariate or bivariate distributions. Section~\ref{sec:pricingUnder_rHW} focuses on derivative pricing; in particular, the pricing equations for options on zero-coupon bonds and swaptions are provided. Numerical results are presented in Section~\ref{sec:numericalExperiments}, where in Section~\ref{sec:evolutionIR}, the swaption implied volatility surfaces for HW and rHW models are compared. Section~\ref{sec:impact_on_IV} analyzes the randomizers' impact on controlling implied volatilities for swaptions. Next, the calibration results are presented in Section~\ref{sec:calibration_MarketData}, and in Section~\ref{sec:pricingBivariate}, the extension to the bivariate randomization case is established. Finally, the convergence results are illustrated in Section~\ref{sec:convergence} and Section~\ref{sec:conclusion} concludes.

\section{The Heath-Jarrow-Morton models, affinity and randomization}
\label{sec:HJM}
Here, we will give an informal presentation of the Heath-Jarrow-Morton (HJM) framework and its relation to arbitrage-free short-rate models. We define the probability space $\left(\Omega,\F(t),\Q\right)$ on which the HJM arbitrage-free condition requires that the instantaneous forward rates, $f(t,T)$, are modelled by the following arbitrage-free SDE:
\begin{equation}
\label{eqn:HJMDynamics}
    \d f(t,T)=\alpha(t,T)\dt+\gamma(t,T)\dW(t),\;\;\;\alpha(t,T)= \gamma(t,T)\int_t^T \gamma(t,z)\dz.
\end{equation}
The instantaneous forward rate, $f(t,T)$, is determined by the volatility driver $\gamma(t,T)$, which can be defined as a deterministic function, stochastic process or by a random variable. The case where $\gamma(t,T)$ is defined as a random variable, thus time-invariant, will be associated with parameter randomization.

We focus here on the short-rate dynamics under the HJM framework. Suppose that $f(0,t)$, $\alpha(t,T)$ and $\gamma(t,T)$ are differentiable in their second argument, with $\int_0^T\left|\frac{\partial}{\partial t}f(0,t)\right|<\infty,$ then the short-rate process under the HJM framework is given by,
\begin{eqnarray}
\d r(t) = \zeta(t)\dt + \gamma(t,t) \d W(t),
\label{eqn:short_rate_dynamics}
\end{eqnarray}
with $$\zeta(t)=\alpha(t,t)+\frac{\partial }{\partial t}f(0,t)+\int_0^t\frac{\partial}{\partial t}\alpha(z,t)\d z+\int_0^t\frac{\partial}{\partial t}\gamma(z,t)\d W(z).$$

A particular class of popular models are the models belonging to the Affine-Diffusion interest rate models. It is shown~\citep{Duffie:2000} that in this class, for process ${{r}}(t)$, the discounted characteristic function is of the following form: \begin{equation}
\label{eqn:ChFCond}
    \phi_{{r}}({{u}};t,T)=\E_t\left[\exp\left(-\int_t^T r(s)\ds+i{u}{{r}}(T)\right)\right]=\exp\left(A({ u};\tau)+{{B}}({ u};\tau){{r}}(t)\right),
\end{equation}
with the expectation under risk-neutral measure $\Q$ for $\tau=T-t$, where $A({u};\tau)$ and ${{ B}}({ u};\tau)$ form the fundamental solution to the corresponding pricing PDE and satisfying the complex-valued ODEs, see the work by Duffie-Pan-Singleton~\citep{Duffie:2000}. The affinity conditions are rather strict as they impose a linear structure on the model, i.e., the model dynamics and its covariance structure needs to be linear in the state variables. The representation in~(\ref{eqn:ChFCond}) is of particular importance as it gives an explicit closed-form expression for the integrated short-rate process and therefore determines the price of a zero-coupon bond (ZCB). Models not leading to an analytic ZCB price are rarely used for pricing purposes. Unfortunately, because of the affinity constraints, the model extensions for some model parameters to be stochastic are minimal. In this article, we focus on relaxing this constraint and letting the model parameters be random while benefiting from the closed-form solutions for the ZCBs and pricing, as presented in~(\ref{eqn:ChFCond}).

We consider a vector $\Theta=[{\vartheta}_1,\dots,{\vartheta}_d]^\T$, with $d\in\mathbb{N}$ representing the number of randomized parameters, where each ${\vartheta}_i$ is a possibly correlated, time-invariant, random variable. A realization of ${\vartheta_i}$ is indicated by $\theta_{i}$, $\vartheta_i(\omega)=\theta_i$, and consequently the realization for ${\Theta}$ is indicated by $\theta=[\theta_{1},\dots,\theta_d]^\T$. Under the RAnD framework, parameters are not driven by a stochastic process but only by a random variable; however, an extension to piece-wise stochastic parameters is also possible (see~\cite{grzelakRAnD} for details).

In a nutshell, the RAnD method relies on quadrature integration and conditional expectation. Therefore, the technique can be applied wherever we deal with an expectation. In principle, randomization can be applied to Equation~(\ref{eqn:ChFCond}), which would provide us with a {\it randomized} ZCB; on the other hand, the randomization can be applied at the valuation level, where the expectation of the expected payoff is computed. We prefer the latter approach to construct the randomized variant of any interest rate model by simply evaluating the model on a particular realization of the model parameters. This will be particularly important when pricing non-standard derivatives.

In the randomized HJM framework, the volatility in~(\ref{eqn:HJMDynamics}) may be dependent on random parameters $\vartheta_1,\dots,\vartheta_d$, i.e., $\gamma(t,T;\{\vartheta_1,\dots,\vartheta_d\}).$
Let us now consider a stochastic model under the HJM framework for derivative pricing with the price denoted as $V(t,r(t;\theta))$ with $\theta$ indicating a vector-valued realization of the model parameter, $\Theta$. $V(t,r(t;\theta))$ may correspond to a model of {Va\v si\v cek}, Hull-White, Black-Karasi{\'n}ski or any other model that belongs to the class of HJM~\cite{BrigoMercurio:2007}. Then, the pricing under randomized parameter space will be given by the following relation:
\begin{eqnarray}
\label{eqn:genericPricing}
V(t,r(t;\Theta)):=\E_t[V(t,r(t;\Theta=\theta))]=\int_{\R^d} V(t,r(t;\Theta=\theta))f_{\Theta}(\theta)\d\theta,
\end{eqnarray}
where the integration takes place over the parameter space $\Theta\in\R^d$ and $f_\Theta(\theta)$ corresponds to the associated $d-$dimensional probability density function.
Equation~(\ref{eqn:genericPricing}) shows that a random model parameter can be considered averaging over possible parameter realizations. From a numerical perspective, the integration in~(\ref{eqn:genericPricing}) is expensive.

The randomization enables for randomizing $d-$dimensional space of parameters; however, for simplicity, we will mainly focus on single or bivariate parameter randomization, $\vartheta_1(\omega)=\theta_1$ and $\vartheta_2(\omega)=\theta_2$, thus also letting for the correlation between stochastic parameters.

Our aim is twofold: we aim to provide numerically efficient methods for the computation of the pricing Equation in~(\ref{eqn:genericPricing}) and to determine the stochastic differential equation for the randomized short-rate processes, $r(t;\vartheta)$ where $\vartheta$ is the chosen random parameter. The insight into the dynamics of the randomized stochastic process will help us categorise the randomized processes and the structure of their drift and volatility coefficients.
\begin{theorem}[Pricing Formula for the Randomized Model]
\label{thm:rHW_generic}
Consider a random variable \(\vartheta\), defined on a finite domain \(D_\vartheta = [a, b]\), with probability density function (PDF) \(f_\vartheta(\theta)\) and cumulative distribution function (CDF) \(F_\vartheta(\theta)\). Assume that the moments up to order \(2N\) are finite, i.e., \(\mathbb{E}[\vartheta^{2N}] < \infty\) for some \(N \in \mathbb{N}\). Let \(V(t, r(t; \theta))\) denote the time \(t\) value of a financial derivative that depends on the short-rate \(r(t; \theta)\) parameterized by \(\theta\), with payoff \(H(T; r(T; \theta))\) at time \(T\). Then, the price of the derivative under the randomized model, \(V(t, r(t; \vartheta))\), is given by:
\begin{equation}
V(t, r(t; \vartheta)) = \int_{[a,b]} V(t, r(t; \theta)) f_\vartheta(\theta) \, \d\theta.
\label{eqn:rHW_Generic}
\end{equation}
Moreover, using an \(N\)-point quadrature approximation (e.g., Gauss quadrature), we have:
\begin{equation}
V(t, r(t; \vartheta)) \approx \sum_{n=1}^N \omega_n V(t, r(t; \theta_n)) + \epsilon_N,
\end{equation}
where the pairs \(\{\omega_n, \theta_n\}_{n=1}^N\) are the quadrature weights and nodes based on the distribution \(f_\vartheta(\theta)\), and \(\epsilon_N\) is the approximation error given by:
\begin{equation}
\epsilon_N = \frac{(b - a)^{2N+1}}{(2N+1)!} \left| \frac{\partial^{2N}}{\partial \theta^{2N}} V(t, r(t; \theta))\Big|_{\theta = \xi} \right|, \quad \text{for some } \xi \in (a, b),
\label{eqn:RAnDError}
\end{equation}
provided that \(V(t, r(t; \theta))\) is \((2N)\) times differentiable with respect to \(\theta\) on \((a, b)\).

\begin{proof}
We start from the standard martingale pricing formula under the risk-neutral measure \(\mathbb{Q}\):
\begin{equation}
V(t, r(t; \vartheta)) = \mathbb{E}_t^{\mathbb{Q}} \left[ \e^{-\int_t^T r(s; \vartheta) \, ds} \, H(T; r(T; \vartheta)) \right].
\end{equation}
Since \(\vartheta\) is a random parameter independent of the Brownian motion driving \(r(t; \theta)\), we can condition on \(\vartheta\):
\begin{equation}
V(t, r(t; \vartheta)) = \mathbb{E}_\vartheta \left[ \mathbb{E}_t^{\mathbb{Q}} \left[ \e^{-\int_t^T r(s; \theta) \, ds} \, H(T; r(T; \theta)) \right] \right] = \int_{a}^{b} V(t, r(t; \theta)) f_\vartheta(\theta) \, \d\theta.
\end{equation}
Here, for each fixed \(\theta\), the inner expectation is the derivative price \(V(t, r(t; \theta))\) under the model parameterized by \(\theta\):
\begin{equation}
V(t, r(t; \theta)) = \mathbb{E}_t^{\mathbb{Q}} \left[ \e^{-\int_t^T r(s; \theta) \, ds} \, H(T; r(T; \theta)) \right].
\end{equation}

Now, for each \(\theta\), we can move from the risk-neutral measure \(\mathbb{Q}\) to the \(T\)-forward measure \(\mathbb{Q}_\theta^T\) associated with \(r(t; \theta)\). The \(T\)-forward measure \(\mathbb{Q}_\theta^T\) depends on \(\theta\) because both the bond price \(P(t, T; \theta)\) and the dynamics of \(r(t; \theta)\) depend on \(\theta\).

Under \(\mathbb{Q}_\theta^T\), we have:
\begin{equation*}
V(t, r(t; \theta)) = P(t, T; \theta) \, \mathbb{E}_t^{\mathbb{Q}_\theta^T} \left[ H(T; r(T; \theta)) \right],
\end{equation*}
where the bond price \(P(t, T; \theta)\) is given by: $
P(t, T; \theta) = \mathbb{E}_t^{\mathbb{Q}} \left[ \e^{-\int_t^T r(s; \theta) \, \ds} \right].$

Therefore, the price under the randomized model becomes:
\begin{equation}
\label{eqn:Vt_integral}
V(t, r(t; \vartheta)) = \int_{a}^{b} P(t, T; \theta) \, \mathbb{E}_t^{\mathbb{Q}_\theta^T} \left[ H(T; r(T; \theta)) \right] f_\vartheta(\theta) \, \d\theta.
\end{equation}
Since both \(P(t, T; \theta)\) and \(\mathbb{E}_t^{\mathbb{Q}_\theta^T} \left[ H(T; r(T; \theta)) \right]\) depend on \(\theta\), we cannot separate the integral further without additional assumptions.

To approximate the integral in Equation~\eqref{eqn:Vt_integral}, we use an \(N\)-point quadrature rule:
\begin{equation*}
V(t, r(t; \vartheta)) \approx \sum_{n=1}^N \omega_n \left[ P(t, T; \theta_n) \, \mathbb{E}_t^{\mathbb{Q}_{\theta_n}^T} \left[ H(T; r(T; \theta_n)) \right] \right] + \epsilon_N,
\end{equation*}
or equivalently,
\begin{equation*}
V(t, r(t; \vartheta)) \approx \sum_{n=1}^N \omega_n V(t, r(t; \theta_n)) + \epsilon_N,
\end{equation*}
where the quadrature nodes \(\theta_n\) and weights \(\omega_n\) are determined based on the distribution \(f_\vartheta(\theta)\), and \(\epsilon_N\) is the approximation error.

The error term \(\epsilon_N\) depends on the quadrature rule and the smoothness of \(V(t, r(t; \theta))\) with respect to \(\theta\). If \(V(t, r(t; \theta))\) is sufficiently smooth, the error \(\epsilon_N\) decreases rapidly with increasing \(N\). Specifically, for Gaussian quadrature, the error can be bounded by:
\begin{equation}
\epsilon_N = \frac{(b - a)^{2N+1}}{(2N+1)!} \left| \frac{\partial^{2N}}{\partial \theta^{2N}} V(t, r(t; \theta))\Big|_{\theta = \xi} \right|, \quad \text{for some } \xi \in (a, b),
\end{equation}
provided that \(V(t, r(t; \theta))\) is \((2N)\) times continuously differentiable with respect to \(\theta\) on \((a, b)\).

This completes the proof.
\end{proof}
\end{theorem}

Theorem~\ref{cor:RAnDChF} gives us an explicit relation between a continuous randomizer $\vartheta$ and the {\it discretization} in terms of pairs $\{w_n,\theta_n\}_{n=1}^N$, leading to a simplification of the integral in~(\ref{eqn:genericPricing}). In this paper, these pairs are based on moments of the randomizer, $\vartheta,$ and, for completeness, the detailed computation procedure is given in~\ref{res:zeta}.

The associated error $\epsilon_N$ in~(\ref{eqn:RAnDError}) can be interpreted as a {\it cost} when moving from the continuous to a discretized random parameter, $\vartheta\rightarrow \{\omega_n,\theta_n\}_{n=1}^N$. To some extent, the pricing with a discrete set of parameter realizations, $\theta_n$, and the associated probabilities, $w_n$, resembles the so-called regime-switch method where a finite set of states is defined. However, from a practical perspective, it is challenging to deal with $N$ pairs of the possible variable states and the associated probabilities, especially in the calibration procedure.

The procedure described in~\ref{res:zeta}, however, simplifies this problem as we can control the quadrature pairs employing the parameters of the randomizer $\vartheta$. To illustrate this process, let us consider a stochastic parameter with a randomizer $\vartheta\sim\mathcal{N}(\mu,\sigma^2)$; thus, the stochastic parameter, $\vartheta$, is driven by two parameters $\mu$ and $\sigma^2$. By application of~\ref{res:zeta}, for any set of parameters $\mu$ and $\sigma^2$, we compute the corresponding pairs $\{\omega_n,\theta_n\}_{n=1}^N$; therefore, in the calibration procedure we will only vary $\mu$, and $\sigma^2$. This procedure drastically reduces the number of associated model parameters, thus facilitating the calibration. We also stress that the set of optimal pairs is solely obtained based on the moments of the randomizing random variable, $\vartheta$, implying that every random variable with, preferably, closed-form moments may be used for randomization. In Table~\ref{Tab:Moments}, a few selected random variables and their moments are tabulated. In Section~\ref{sec:calibration_MarketData}, we will also classify the randomizers, $\vartheta$, for which the computations can be greatly reduced or even tabulated.

To emphasize the flexibility of the RAnD method, we focus on the application in the Hull-White model, where we will apply randomization to each model parameter and consider an extension to a bivariate case. In the following sections, we consider a finite number of realizations of $\vartheta$. We will denote them as $\theta_1,\dots,\theta_N$, for some $N\in\mathbb{N}$. These ``specific'' realizations we will interchangeably call either ``collocation''~\citep{scmc2019} or ``quadrature'' points. Finally, by $\vartheta(\hat a,\hat b)$, we denote that $\vartheta$ is a random variable driven by parameters: $\hat a$, $\hat b$.

\subsection{Arbitrage-free conditions under the RAnD method}
\label{sec:arbitrageHJM}
Before we analyze the specifics of the randomized Hull-White (rHW) model, let us review the implications of randomization on the pricing, under the affine-diffusion framework, of ZCB and discuss arbitrage-free aspects of the RAnD method. By setting $u=0$ in~(\ref{eqn:ChFCond}), the risk-free pricing formula for a randomized ZCB, $P(t, T;\vartheta)$ is given by:
\begin{eqnarray}
\label{eqn:ZCB}
P(t,T;\vartheta)=\int_{[a,b]} P(t,T;\theta)\d F_{\vartheta}(\theta)=\int_{[a,b]} \e^{A(\tau;\theta)+{{B}}(\tau;\theta){{r}}(t;\theta)}\d F_{\vartheta}(\theta),\;\;a<b,
\end{eqnarray}
with $A(\tau;\theta):=A(0,\tau;\theta)$ and $B(\tau;\theta):=A(0,\tau;\theta)$ in~(\ref{eqn:ChFCond}) for $\tau=T-t,$ and $r(t;\theta)$ indicates a short-rate model with constant parameter realizations $\theta.$ By application of Theorem~\ref{thm:rHW_generic}, the randomized ZCB is known explicitly, and it is presented in Corollary~\ref{cor:RAnDChF}.

\begin{cor}[ZCBs under Randomized Affine Jump Diffusion Processes]
\label{cor:RAnDChF}
Let ${r}(t;\theta)$ represent an affine short-rate process with some constant parameter $\theta$. Assuming that the corresponding ChF, $\phi_{r(T)|\vartheta=\theta}(\cdot)$, is well defined and $2N$ times differentiable w.r.t. $\theta$, the randomized ZCB is given by,
\begin{eqnarray}
P(t,T;\vartheta)&=&\sum_{n=1}^{N}\omega_nP(t,T;\theta_n)+\epsilon_N=\sum_{n=1}^{N}\omega_n\e^{A(\tau;\theta_n)+{{ B}}(\tau;\theta_n)r(t;\theta_n)}+\epsilon_N\nonumber\\
\label{eqn:randChF_ZCB}
&=:&P(t,T;\{\theta_n\}_{n=1}^N)+\epsilon_N,
\end{eqnarray}
with ${A}(\tau;\theta_n)$, ${{ B}}(\tau;\theta_n)$ being the real-valued functions obtained from Riccati-type of ODEs available for affine models~\cite{OosterleeGrzelakBook}.
\begin{proof}
The proof is a consequence of Theorem~\ref{thm:rHW_generic}.
\end{proof}
\end{cor}
Representation~(\ref{eqn:randChF_ZCB}) illustrates that the ZCB, $P(t,T;\vartheta)$, randomized with stochastic parameter $\vartheta$ can be expressed as a weighted sum of the ZCBs evaluated at a given realization of $\vartheta$ and where $\omega_1+\dots+\omega_N=1$. Note that under the HJM framework, each ZCB, $P(t,T;\theta_n)$ in~(\ref{eqn:randChF_ZCB}) is arbitrage-free; therefore, the convex combination is too. This can be shown by checking whether $P(S,T;\{\theta_n\}_{n=1}^N)/M(S)$ is indeed a martingale for some $t<S<T$:
\begin{eqnarray*}
\E_t\left[\frac{P(S,T;\{\theta_n\}_{n=1}^N)}{M(S)}\right]=\E_t\left[\frac{\sum_{n=1}^{N}\omega_nP(S,T;\theta_n)}{M(S)}\right]=\sum_{n=1}^{N}\omega_n\E_t\left[\frac{P(S,T;\theta_n)}{M(S)}\right],
\end{eqnarray*}
since every discounted ZCB, $P(S,T;\theta_n)/M(S),$ is a martingale under the $\Q$ measure we have,
\begin{eqnarray*}
\E_t\left[\frac{P(S,T;\{\theta_n\}_{n=1}^N)}{M(S)}\right]=\sum_{n=1}^{N}\omega_n\frac{P(t,T;\theta_n)}{M(t)}=\frac{P(t,T;\{\theta_n\}_{n=1}^N)}{M(t)}.
\end{eqnarray*}
This result shows that the application of the RAnD method enables the transition from a continuous random variable to a discrete one; the cost of this transition is expressed by $\epsilon_N$. Moreover, as shown above, the error $\epsilon_N$ in~(\ref{eqn:randChF_ZCB}) does not impact the arbitrage conditions-the pricing under the RAnD method is simply a {\it weighted average} of arbitrage-free prices. However, from a practical perspective, keeping this error as small as possible is still essential. This is particularly important when calibrating the randomized model to market data, i.e., the calibration procedure is associated with varying parameters of the variable $\vartheta$. It is, therefore, important that the discretized version resembles the continuous version as closely as possible. In the numerical section, Section~\ref{sec:convergence}, we will investigate the convergence aspects of different randomizers and an optimal number of expansion terms, $N$.


\subsection{The randomized Hull-White (rHW) short-rate model}
\label{sec:randomizedHW}
This section provides a specification leading us to the randomized version of the famous HW model~\citep{HullWhite:1990}-a single-factor, no-arbitrage yield curve model, in which an extended Ornstein-Uhlenbeck drives the short-term interest rate mean-reverting process. Under the HJM framework and the arbitrage-free condition for the drift in~(\ref{eqn:HJMDynamics}), the rHW model is specified by:
\begin{equation}
 \label{eqn:HJMHW}
\gamma(t,T)=\eta\cdot \e^{-\lambda(T-t)},\;\;\;t<T,
\end{equation}
where we consider three different randomization cases: the randomization of the volatility parameter, $\eta$, the mean-reversion, $\lambda$, or the randomization of both parameters using bivariate distribution:
\begin{equation}
\label{eqn:HW_RAnD_Parms}
\eta\stackrel{\d}{=}\vartheta_1,\;\;\text{or}\;\;\; \lambda\stackrel{\d}{=}\vartheta_2,\;\;\;\text{or}\;\;\; \lambda|\eta\stackrel{\d}{=}\vartheta_2|\vartheta_1.
\end{equation}
To simplify the notation, we consider one parameter $\theta\in\{\eta,\lambda\}$ with the corresponding random variable $\vartheta$. Particular choices of the parameters will be given explicitly.

Given the HJM volatility in~(\ref{eqn:HJMHW}), under the risk-free measure, the dynamics of the HW model read:
\begin{equation}
\label{eqn:HW_SDE}
    \d {r}(t)=\lambda(\psi(t)-{r}(t))\dt + \eta\dW(t),\;\;\;r_0\equiv f(0,0),
\end{equation}
with \begin{equation}
\label{eqn:Psi_f_0_t}
\psi(t)=f(0,t)+\frac{1}{\lambda}f(0,t)+\frac{\eta^2}{2\lambda^2}\left(1-\e^{-2\lambda t}\right),\;\;\;f(0,t)=-\frac{\partial \log P(0,t)}{\partial t},
\end{equation}
where $\psi(t)$ is a time-dependent drift term, which ensures the fit of the calibration to a yield curve observed in the market, $W(t)$ is the Brownian motion under measure $\Q$, and $f(0,t)$ indicates the instantaneous forward rate computed from the yield curve, defined in terms of ZCBs, $P(0,t)$. Parameter $\eta$ determines the overall level of the volatility, and $\lambda$ is the reversion rate parameter. A large value of $\lambda$ causes short-term rate movements to dampen out rapidly, reducing the long-term volatility. Because of this interdependence between the model parameters, often, in practical applications, $\lambda$ is fixed and set to a constant. In contrast, $\eta$ is often set to be piece-wise constant and calibrated such that the ATM volatilities are calibrated. We will show that under the RAnD method, such a strategy is inadequate.

The short-rate $r(t)$ in~(\ref{eqn:HW_SDE}) is thus {\it normally distributed} with $r(t)\sim\mathcal{N}(\mu_{r(t)},\sigma^2_{r(t)}),$ with
\begin{eqnarray}
\label{eqn:r_t_distribution}
    \mu_{r(t)}:=r_0\e^{-\lambda
t}+\lambda\int_0^t{\psi(z)\e^{-\lambda(t-z)} \dz},\;\;\;\sigma^2_{r(t)}:=\frac{\eta^2}{2\lambda}\left(
1-\e^{-2\lambda t}\right).
\end{eqnarray}
Moreover, for
$\psi(t)$ constant, i.e., $\psi(t)\equiv\psi$ (in this case we deal with the \index{Va\v si\v cek model}{\em Va\v si\v cek model}~\citep{Vasicek:1977}), we have $\lim_{t\rightarrow\infty}\E_{t_0}\left[r(t)\right]=\psi.$ This
means that the first moment of the process converges
to the mean-reverting level $\psi$, for large values of $t$.

As a first step, we check the impact of randomization on the probability density function, PDF, of the interest rate process $r(t)$ in~(\ref{eqn:HW_SDE}). Since the HW model belongs to the affine class of processes, we can benefit from the available ChF. Given a stochastic parameter ${\bf \vartheta}$, the ChF is given by:
\begin{eqnarray*}
\phi_{{ r}}({ u};t,T):=\E_t\Big[\e^{-\int_t^T r(s)\ds+i{ u}{r}(T)}\Big]=\E_t\left[\E_t\Big[\e^{-\int_t^T r(s)\ds+i{ u}{ r}(T)}\big|{\vartheta}={\theta}\Big]\right].
\end{eqnarray*}
By application of the quadrature rule in Theorem~\ref{thm:rHW_generic} and~\ref{res:zeta}, we find:
\begin{eqnarray}
\label{eqn:ChFIntegral}
\phi_{{ r}}({ u};t,T)=\int_{[a,b]} \phi_{{r}|{\vartheta}=\theta}({ u};t,T)\d F_{\vartheta}(\theta)=\sum_{n=1}^N\omega_n\phi_{{ {r}(T)}|\vartheta=\theta_n}({ u};t,T) + \epsilon_N^F,
\end{eqnarray}
and by utilizing the Fourier transform, the PDF of the rHW model reads,
\begin{eqnarray}
\nonumber
f_{{{r}(T)}}(x)=\frac{1}{2\pi}\int_{\R}\e^{-iux}\sum_{n=1}^{N}\omega_n\phi_{{ {r}}|\vartheta=\theta_n}({ u};t,T)\d u+\epsilon_N^F=\sum_{n=1}^{N}\omega_nf_{{ r(T;\theta_n)}}(x)+\epsilon_N^F,
\label{eqn:HW_ranDensity}
\end{eqnarray}
where $r(t)|\vartheta=\theta_n$ indicates the HW model with a particular realization, $\theta_n$, of the randomized variable. The representation above presents the relation between the densities of the rHW process, $r(t)$, as a convex combination of the HW processes, $r(t)|\theta_n$ with a particular parameter $\theta_n$. We also highlight that the error $\epsilon_N^F$ in~(\ref{eqn:ChFIntegral}) may differ from the quadrature error in~(\ref{eqn:RAnDError}).

Because of different randomization choices, we distinguish the following randomization types and their associated error:
\begin{equation}
\label{eqn:PDF_rand}
    f_{{{r}(T)}}(y)=\sum_{n=1}^{N}\omega_nf_{{ r(T;\theta_n)}}(y)+\epsilon_N^F=:\left\{\begin{array}{ccc}
    f_{\overline{r}(T)}(y)+\overline\epsilon_N,\;\;\;\text{for}\;\;\;\eta\stackrel{\d}{=}\vartheta,\\
    f_{\widetilde{r}(T)}(y)+\widetilde\epsilon_N,\;\;\;\text{for}\;\;\;\lambda\stackrel{\d}{=}\vartheta,\\
    f_{\widehat{r}(T)}(y)+\widehat\epsilon_N,\;\;\;\text{for}\;\;\;\eta\stackrel{\d}{=}\vartheta_1\;\;\&\;\;\lambda|\eta\stackrel{\d}{=}\vartheta_2.
    \end{array}\right.
\end{equation}
Under the RAnD method, the PDF, $f_{{{r}(T)}}(y)$ in~(\ref{eqn:PDF_rand}) is given as a linear combination of normal densities that depend on different parameter realizations $\theta_n$, resembling a similar problem as presented in~\cite{brigo2002lognormal}.

In the next section, we associate the PDF given in~(\ref{eqn:PDF_rand}) and  find the corresponding SDE for a 1D stochastic representation.
\subsection{Dynamics of the rHW model}
\label{sec:dynamics_rHW}
The combination of Equations~(\ref{eqn:HW_RAnD_Parms}) and~(\ref{eqn:HW_SDE}) shows us that the simulation of  the randomized short-rate models is explicit using the Monte Carlo technique. However, by an explicit form of the corresponding SDE, we will get more insight into the randomization and its impact on the dynamics of the driving process.

As a start, under the rHW model, consider the randomization of the volatility parameter, $\theta:=\eta$, with $\vartheta(\omega)=\theta$, and where the corresponding quadrature pairs are given by $\{\omega_n,\eta_n\}_{n=1}^N$. For such a setting, under each realization $\eta_n$, we have the associated HW model, $\overline{r}_n(t):=\overline{r}_n(t;\eta_n)$, with the dynamics given by:
\begin{eqnarray}
\label{eqn:HW_eta_n}
\d \overline{r}_n(t)=\lambda(\overline\psi_n(t)-\overline{r}_n(t))\dt + \eta_n\dW(t),\;\;\;\overline{r}_n(t_0)=f(0,0),\;\;\;n=1,\dots,N,
\end{eqnarray}
with common Brownian motion, $W(t)$, and initial value $\overline{r}_n(t_0)=f(0,0)$ in~(\ref{eqn:Psi_f_0_t}), where $\lambda$ is constant and equal for all the $\overline{r}_n(t)$, and where $\overline\psi_n(t)$ is defined in~(\ref{eqn:Psi_f_0_t}) and is given as a function of $\lambda$ and $\eta_n$.

Given the sequence of HW model processes, $\overline{r}_1(t),\dots,\overline{r}_N(t)$, in~(\ref{eqn:HW_eta_n}) and the probability density relation in~(\ref{eqn:HW_ranDensity}), we consider the problem of finding the corresponding SDE for the rHW process, $\overline{r}(t)$. Formally, we seek an SDE, with the solution given in~(\ref{eqn:PDF_rand}) and where each of the constituent processes, $\overline{r}_n(t)$, is driven by~(\ref{eqn:HW_eta_n}).
Thus, we consider the following process,
\begin{eqnarray}
\label{eqn:HW_local}
\d \overline{r}(t)=\overline\lambda(t,\overline{r}(t))\dt + \overline{\eta}(t,\overline{r}(t))\dW(t),\;\;\;\overline{r}(t_0)=f(0,0),
\end{eqnarray}
with some state-dependent drift, $\overline\lambda(t,\overline{r}(t))$, and volatility, $\overline{\eta}(t,\overline{r}(t))$, and where Brownian motion $W(t)$ is as in~(\ref{eqn:HW_eta_n}).
As indicated in~\cite{brigo2008general}, the problem of finding an SDE when marginal distributions and the corresponding weights are given is the reverse of finding the marginal density function of the solution of an SDE when the coefficients are known. Moreover, the problem of a mixture of normal or lognormal processes is known~\cite{brigo2008general}. Here, in Proposition~\ref{prop:rand_eta}, we adapt those techniques and provide an explicit form for the SDE in~(\ref{eqn:HW_local}).
\begin{prop}[Local volatility process for the HW model with randomized volatility parameter, $\eta$]
\label{prop:rand_eta}
Let us assume a sequence of positive constants $\eta_n$, $n=1,\dots,N.$ Then, the SDE
\begin{equation}
\label{eqn:HW_localVol}
\d \overline{r}(t)=\overline\lambda(t,\overline{r}(t))\dt + \overline{\eta}(t,\overline{r}(t))\dW(t),\;\;\;\overline{r}(t_0)=f(0,0),
\end{equation}
with
\begin{eqnarray}
\overline\lambda(t,y)=\sum_{n=1}^{N}\overline{\Lambda}_n(t,y)\lambda(\overline\psi_n(t)-y),\;\;\;\overline{\eta}^2(t,y)=\sum_{n=1}^{N}\eta_n^2\overline\Lambda_n(t,y),
\end{eqnarray}
where:
\[\overline\Lambda_n(t,y)=\frac{\omega_nf_{\overline{r}(t;\eta_n)}(y)}{\sum_{n=1}^{N}\omega_nf_{\overline{r}(t;\eta_n)}(y)},\]
has a strong solution whose marginal density is given by the mixture of normal probability density functions:
\begin{eqnarray}
\label{eqn:HW_rand_eta_density}
f_{{\overline{r}(t)}}(y)=\sum_{n=1}^{N}\omega_nf_{\overline{r}(t;\eta_n)}(y),
\end{eqnarray}
where $\sum_{n=1}^N\omega_n=1$ for $\omega_n\geq0$, $n=1,\dots,N$ with $f_{\overline{r}(t;\eta_n)}(x)$ the PDF of the HW model with dynamics, given by:
\begin{equation}
\label{eqn:HW_individual_asset}
\d \overline{r}_n(t)=\lambda(\overline\psi_n(t)-\overline{r}_n(t))\dt + \eta_n\dW(t),\;\;\;\overline{r}_n(t_0)=f(0,0),
\end{equation}
where $\overline{r}_n(t):=\overline{r}_n(t;\eta_n)$ with $\overline\psi_n(t)=f(0,t)+\frac{1}{\lambda}f(0,t)+\frac{\eta_n^2}{2\lambda^2}\left(1-\e^{-2\lambda t}\right).$
\begin{proof}
The problem we address is the derivation of the drift, $\overline\lambda(t,y)$, and the volatility function, $\overline\eta(t,\overline{r}(t))$, in the SDE:
\begin{equation}
    \label{eqn:HW_eta}
    \d \overline{r}(t)=\overline\lambda(t,\overline{r}(t))\dt + \overline\eta(t,\overline{r}(t))\dW(t),
\end{equation}
such that \begin{equation}
\label{eqn:density_r_T}
f_{{\overline{r}(t)}}(y)=\sum_{n=1}^{N}\omega_nf_{\overline{r}(t;\eta_n)}(y),
\end{equation}
where $f_{\overline{r}(t;\eta)}(y)$ is the PDF of the Hull-White process in~(\ref{eqn:HW_eta_n}), with parameter $\eta$. Under the HW model dynamics, the so-called linear-growth condition holds, i.e., $\eta_n^2\leq C_n(1+y^2)$ for $n=1,\dots,N.$ Assuming that the same non-explosion condition holds for~(\ref{eqn:HW_eta}), i.e., $\overline\eta^2(t,y)\leq C(1+y^2),$ uniformly in $t,$ we start by deriving the
Fokker-Planck equation for $\overline{r}(t)$:
\begin{eqnarray}
\label{eqn:FK_r_T}
\frac{\partial }{\partial t}f_{\overline{r}(t)}(y)=\frac12\frac{\partial^2 }{\partial y^2}\left(\overline\eta^2(t,y))f_{\overline{r}(t)}(y)\right)-\frac{\partial }{\partial y}\overline\lambda(t,y)f_{\overline{r}(t)}(y),
\end{eqnarray}
while for each individual interest rate process, $r_n(t)$, in~(\ref{eqn:HW_individual_asset}) we have:
\begin{eqnarray}
\label{eqn:SDE_individual_r_t}
\frac{\partial }{\partial t}f_{\overline{r}(t;\eta_n)}(y)=\frac12\frac{\partial^2 }{\partial y^2}\left(\eta_n^2f_{\overline{r}(t;\eta_n)}(y)\right)-\frac{\partial }{\partial y}\left(\lambda(\overline\psi_n(t)-y)\right)f_{\overline{r}(t;\eta_n)}(y).
\end{eqnarray}
After substituting~(\ref{eqn:density_r_T}) into~(\ref{eqn:FK_r_T}):
\begin{eqnarray*}
\frac{\partial }{\partial t}\sum_{n=1}^{N}\omega_nf_{\overline{r}(t;\eta_n)}(y)=\frac12\frac{\partial^2 }{\partial y^2}\left[\overline\eta^2(t,y)\sum_{n=1}^{N}\omega_nf_{\overline{r}(t;\eta_n)}(y)\right]-\frac{\partial }{\partial y}\overline{\lambda}(t,y)\sum_{n=1}^{N}\omega_nf_{\overline{r}(t;\eta_n)}(y).
\end{eqnarray*}
Due to the linearity of the derivative operator, we find,
\begin{eqnarray*}
\sum_{n=1}^{N}\omega_n\frac{\partial }{\partial t}f_{\overline{r}(t;\eta_n)}(y)=\frac12\frac{\partial^2 }{\partial y^2}\left[\overline\eta^2(t,y)\sum_{n=1}^{N}\omega_nf_{\overline{r}(t;\eta_n)}(y)\right]-\frac{\partial }{\partial y}\overline\lambda(t,y)\sum_{n=1}^{N}\omega_nf_{\overline{r}(t;\eta_n)}(y).
\end{eqnarray*}
By substitution of~(\ref{eqn:SDE_individual_r_t}) and operator linearity,
\begin{eqnarray*}
\small
\frac{\partial^2 }{\partial y^2}\sum_{n=1}^{N}\omega_n\frac12\eta_n^2f_{\overline{r}(t;\eta_n)}(y)-\sum_{n=1}^{N}\omega_n\frac{\partial }{\partial y}\lambda(\overline\psi_n(t)-y)f_{\overline{r}(t;\eta_n)}(y)=\\\frac12\frac{\partial^2 }{\partial y^2}\left[\overline\eta^2(t,y)\sum_{n=1}^{N}\omega_nf_{\overline{r}(t;\eta_n)}(y)\right]-\frac{\partial }{\partial y}\overline\lambda(t,y)\sum_{n=1}^{N}\omega_nf_{\overline{r}(t;\eta_n)}(y).
\end{eqnarray*}
Finally, by matching the appropriate terms and integration, we find:
\begin{eqnarray*}
\sum_{n=1}^{N}\omega_n\eta_n^2f_{\overline{r}(t;\eta_n)}(y)&=&\overline\eta^2(t,y)\sum_{n=1}^{N}\omega_nf_{\overline{r}(t;\eta_n)}(y)+C_1(t)y+C_2(t),\\
\sum_{n=1}^{N}\omega_n\lambda(\overline\psi_n(t)-y)f_{\overline{r}(t;\eta_n)}(y)&=&\overline\lambda(t,y)\sum_{n=1}^{N}\omega_nf_{\overline{r}(t;\eta_n)}(y)+C_3(t),
\end{eqnarray*}
for some time-dependent functions $C_1(t)$, $C_2(t)$ and $C_3(t).$
Since the LHS and RHS of the equations above need to satisfy uniform convergence requirements implying that for $y\rightarrow\infty$, they converge to $0$, the following needs to hold: $C_1(t)=C_2(t)=C_3(t)=0$, $\forall t.$ Therefore, the expressions for $\overline\lambda(t,y)$ and $\overline\eta(t,y)$ read:
\begin{eqnarray}
\overline\eta^2(t,y)=\frac{\sum_{n=1}^{N}\omega_n\eta_n^2f_{\overline{r}(t;\eta_n)}(y)}{\sum_{n=1}^{N}\omega_nf_{\overline{r}(t;\eta_n)}(y)},\;\;\;\;\;\overline\lambda(t,y)=\frac{\sum_{n=1}^{N}\omega_n\overline\lambda(\overline\psi_n(t)-y)f_{\overline{r}(t;\eta_n)}(y)}{\sum_{n=1}^{N}\omega_nf_{\overline{r}(t;\eta_n)}(y)}.
\end{eqnarray}
By setting
\begin{equation}
\overline\Lambda_n(t,y)=\frac{\omega_nf_{\overline{r}(t;\eta_n)}(y)}{\sum_{n=1}^{N}\omega_nf_{\overline{r}(t;\eta_n)}(y)}\;\;\;\text{for}\;\;\;n=1,\dots,N,
\end{equation}
we can write
\begin{eqnarray}
\overline\lambda(t,y)=\sum_{n=1}^{N}\overline{\Lambda}_n(t,y)\lambda(\overline\psi_n(t)-y),\;\;\;\text{and}\;\;\;\overline{\eta}^2(t,y)=\sum_{n=1}^{N}\eta_n^2\overline\Lambda_n(t,y).
\end{eqnarray}
Finally, by taking $\eta_{*}:=\max_{i=1,\dots,N}\eta_n$ and since $\sum_{n=1}^N\overline{\Lambda}_n(t,y)=1,$ $\forall y,$ we have:
\begin{eqnarray*}
\overline\eta^2(t,y)=\sum_{n=1}^{N}\eta_n^2\overline\Lambda_n(t,y)\leq \sum_{n=1}^{N}\eta_{*}^2\overline\Lambda_n(t,y)=\eta_*^2=C.
\end{eqnarray*}
Since the volatility parameter $\overline\eta^2(t,y)$ is bounded by a constant, the uniform convergence criterion is satisfied. The uniqueness of the strong solution follows from Theorem 12.1 in~\cite{rogers_williams_2000} while in~\cite{brigo2008general} (Theorem 2.1), a proof for a generic case for a normal mixture is provided.
\end{proof}
\end{prop}
Proposition~\ref{prop:rand_eta} illustrates that under the randomized volatility parameter for the HW model, the normal mixture dynamics resemble a one-dimensional local-volatility-type diffusion process. The local volatility function is expressed as a weighted volatility squared, $\eta_n^2$, of the constituent processes, $\overline{r}_n(t)$, and where the weights are functions of the quadrature coefficients, $\omega_n$, and the corresponding PDFs, $f_{\overline{r}(t;\eta_n)}(y)$.

Following the same strategy, we derive the dynamics of a process of the rHW model with the randomized mean-reversion parameter, $\lambda$. Proposition~\ref{prop:rand_lambda} provides the details.
\begin{prop}[Dynamics of the HW model with randomized mean-reversion parameter, $\lambda$] \label{prop:rand_lambda}
Let us assume a sequence of positive constants $\lambda_m$, $m=1,\dots,M$, then the SDE
\begin{equation}
\label{eqn:HW_localVol_lambda}
\d \widetilde{r}(t)=\widetilde\lambda(t,\widetilde{r}(t))\dt + \widetilde{\eta}(t,\widetilde{r}(t))\dW(t),\;\;\;\widetilde{r}(t_0)=f(0,0),
\end{equation}
with $\widetilde\eta(t,y)=\eta$, and
\begin{eqnarray}
\widetilde\lambda(t,y)=\sum_{m=1}^M\widetilde\Lambda_m(t,y)\lambda_m(\psi_m(t)-y),\;\;\;\text{where}\;\;\;\widetilde\Lambda_m(t,y)=\frac{\varpi_mf_{\widetilde{r}(t;\lambda_m)}(y)}{\sum_{m=1}^{M}\varpi_mf_{\widetilde{r}(t;\lambda_m)}(y)},
\end{eqnarray}
has a strong solution whose marginal density is given by the mixture of normal probability density functions:
\begin{eqnarray}
\label{eqn:HW_rand_lambda_density}
f_{{\widetilde{r}(t)}}(x)=\sum_{m=1}^{M}\varpi_mf_{\widetilde{r}(t;\lambda_m)}(x),
\end{eqnarray}
where $\sum_{m=1}^M\varpi_m=1$ for $\varpi_m\geq0$, $m=1,\dots,M$, with $f_{\widetilde{r}(t;\lambda_m)}(x)$ is the PDF the HW model whose dynamics is given by:
\[\d \widetilde{r}_m(t)=\lambda_m(\psi_m(t)-\widetilde{r}_m(t))\dt + \eta\dW(t),\;\;\;\widetilde{r}_m(t_0)=f(0,0),\]
where $\widetilde{r}_n(t):=\widetilde{r}_n(t;\lambda_n)$ with $\psi_m(t)=f(0,t)+\frac{1}{\lambda_m}f(0,t)+\frac{\eta^2}{2\lambda_m^2}\left(1-\e^{-2\lambda_m t}\right).$
\begin{proof}
The proof is analogous to Proposition~\ref{prop:rand_eta}.
\end{proof}
\end{prop}

With the rHW model, the randomization of either model parameters enables us to determine the corresponding 1D SDE. For the random volatility parameter, $\eta$, we derived a local-volatility type short-rate process; however, in the case of random $\lambda$, the corresponding 1D process has a different structure, i.e., the volatility coefficient, $\widetilde\eta(t,y)=\eta$, in~(\ref{eqn:HW_localVol_lambda}), stays constant. When analyzing the dynamics of the corresponding ZCB (see Corollary~\ref{cor:RAnDChF}), which is a tradable asset, the dynamics will resemble a local-volatility type process. To derive the corresponding SDE, one needs to follow the same strategy as in Propositions~\ref{prop:rand_eta} and~\ref{prop:rand_lambda} with the process for the ZCB, instead of the short-rate process.

In the final part of this section, the case of joint randomization of both HW model parameters will be discussed.

\subsubsection{Dynamics of the rHW model with bivariate distribution for $\lambda$ and $\eta$.}
\label{sec:rHW_bivariateCase}
An extension of the RAnD method is to consider both HW model parameters random and follow a bivariate distribution. Such an extension may benefit from the interconnection between model parameters and possibly the correlation coefficient; however, it would require that the conditional moments are known explicitly. This is troublesome because only for a few random distributions the moment functions are known in the closed form.  However, if we stay, for example, within the Gaussian world, such an extension to the 2D case is possible.
\begin{cor}[Random parameters with bivariate distribution]
\label{cor:bivariate}
Under a bivariate distribution $\Theta=[\vartheta_1,\vartheta_2]$ with $\zeta(\vartheta_1)=\{\omega_{1,n},\theta_{1,n}\}_{n=1}^N$ and conditioned on $\zeta(\vartheta_2|\vartheta_1)=\{\omega_{2,m},\theta_{2,m}\}_{m=1}^M$, the randomized ChF is given by:
\begin{eqnarray}
\phi_{{ X}}({ u};t,T)=\sum_{n=1}^{N}\omega_{n}\sum_{m=1}^{M}\omega_{m}\phi_{{ X}|\vartheta_1=\theta_{n},\vartheta_2=\theta_{m}}({ u};t,T)+\epsilon_{N,M}^b,
\end{eqnarray}
where $N$ and $M$ indicate the number of expansion terms for $\vartheta_1$ and $\vartheta_2|\vartheta_1$ respectively, $\vartheta_2|\vartheta_1$ indicates a conditional random variable, $\epsilon_{N,M}^b$ is the corresponding aggregated error, and the remaining specification follows Theorem~\ref{cor:RAnDChF}.
\end{cor}
The PDF computation for the bivariate case requires the sequential computation of the associated weights and the corresponding points, which can be established utilizing Propositions~\ref{prop:rand_eta} and~\ref{prop:rand_lambda}. In the first iteration step, we compute the grid points associated with the volatility parameter, $\eta$, and for each realization $\eta_n$, we establish the corresponding conditional PDF. By summing over all possible pairs, the unconditional PDF for the rHW model can be computed,
\begin{eqnarray}
\label{eqn:PDF_bivariate}
f_{r(T;\eta_n)}(x)=\sum_{m=1}^{M}\varpi_{m,n}f_{{ r(T;\eta_n,\lambda_m)}}(x),\;\;\;f_{{r(T)}}(x)=\sum_{n=1}^{N}\sum_{m=1}^{M}\omega_n\varpi_{m,n}f_{{ r(T;\eta_n,\lambda_m)}}(x),
\end{eqnarray}
where $\varpi_{m,n}$ is defined in~(\ref{eqn:HW_rand_lambda_density}), $\omega_n$ is given in~(\ref{eqn:HW_rand_eta_density}) and $f_{{ r(T;\eta_n,\lambda_m)}}(x)$ indicates the PDF of the HW model with parameters $\eta_n$ and $\lambda_m$, and is defined in~(\ref{eqn:r_t_distribution}).
Proposition~\ref{prop:rand_eta_and_lambda} provides the dynamics of the associated rHW model.
\begin{prop}[Local volatility process for the HW model with randomized parameters]
\label{prop:rand_eta_and_lambda}
Let us assume a sequence of positive constants $\eta_n$, $n=1,\dots,N$, and $\lambda_{m,n}$, $m=1,\dots,M$, then the SDE:
\begin{equation}
\label{eqn:HW_localVol}
\d \widehat{r}(t)=\widehat\lambda(t,\widehat{r}(t))\dt + \widehat{\eta}(t,\widehat{r}(t))\dW(t),\;\;\;\widehat{r}(t_0)=f(0,0),
\end{equation}
with
\begin{eqnarray*}
\widehat\eta^2(t,\widehat{r}(t))&=&\sum_{n=1}^N\sum_{m=1}^M\widehat\Lambda_{n,m}(t,y)\eta_n^2,\;\;\;\widehat\lambda(t,\widehat{r}(t))=\sum_{n=1}^{N}\sum_{m=1}^{M}\widehat\Lambda_{n,m}(t,y)\lambda_{m,n}(\psi_{m,n}(t)-y),
\end{eqnarray*}
where \[\widehat\Lambda_{n,m}(t,y)=\frac{\omega_n\varpi_{m,n}f_{{ r(t;\eta_n,\lambda_m)}}(y)}{\sum_{n=1}^{N}\sum_{m=1}^{M}\omega_n\varpi_{m,n}f_{r(t;\eta_n,\lambda_{m,n})}(y)},\]
has a strong solution whose marginal density is given by the following mixture of normal probability density functions:
\begin{eqnarray*}
f_{{\widehat{r}(t)}}(y)=\sum_{n=1}^{N}\sum_{m=1}^{M}\omega_n\varpi_{m,n}f_{{ r(t;\eta_n,\lambda_{m,n})}}(y),
\end{eqnarray*}
where $\sum_{n=1}^N\omega_n=1$, $\sum_{m=1}^M\varpi_{m,n}=1$ for $\omega_n,\varpi_{m,n}\geq0$, $n=1,\dots,N$, $m=1,\dots,M$ with $f_{r(t;\eta_n,\lambda_{m,n})}(y)$ being the PDF of the HW model whose dynamics are given by:
\[\d r_{m,n}(t)=\lambda_{m,n}(\psi_{m,n}(t)-r_{m,n}(t))\dt + \eta_{n}\dW(t),\;\;\;r_{m,n}(t_0)=f(0,0),\]
and  $r_{n,m}(t):=r(t;\eta_n,\lambda_m)$  with $\psi_{m,n}(t)=f(0,t)+\frac{1}{\lambda_{m,n}}f(0,t)+\frac{\eta_{n}^2}{2\lambda_{m,n}^2}\left(1-\e^{-2\lambda_{m,n} t}\right).$
\begin{proof}
The proof is analogous to Proposition~\ref{prop:rand_eta}.
\end{proof}
\end{prop}
The concept of multivariate model parameters can be extended further. In the case of the HW model with piece-wise constant parameters, it is possible to randomize each piece-wise element, to mimic the parameter controlled by the stochastic process. Unfortunately, as the number of ranges increases, the number of terms in the summation in~(\ref{eqn:PDF_bivariate}) will grow exponentially.

The impact of bivariate model parameters on implied volatilities will be analyzed further in Section~\ref{sec:numericalExperiments}.
\section{Pricing under the randomized Hull-White (rHW) model}
\label{sec:pricingUnder_rHW}
This section focuses on the pricing under the rHW model. The presented pricing equations utilize Theorem~\ref{thm:rHW_generic}. Throughout the section, we denote by $V_{\HW}(t,r(t;\theta))$ with $\theta\in\{\eta,\lambda\}$, the Hull-White price of a derivative depending on the short-rate $r(t;\theta)$ with parameters $\lambda$ and $\eta$, by $V_{\rHW}(t,r(t;\vartheta))$ we denote the pricing under the rHW model.
\subsection{Swaptions under rHW model}
While in Equation~(\ref{eqn:rHW_Generic}), there is a relationship between HW and rHW prices, here explicit pricing formulas will be given. The valuation of swaption contracts in the rHW model can be carried out at the same complexity level as in the standard HW model.

The critical ingredient for pricing swaptions under the rHW model is an analytical expression for the options of the {\it pure} ZCB. Lemma~\ref{lem:optionsZCB} provides the pricing equations under the rHW model.
\begin{lem}[Pricing of options on a ZCB under the rHW model]
\label{lem:optionsZCB}
Under the randomized HW model parameters, $\theta:=\eta$ or $\theta:=\lambda$, with $\vartheta(\omega)=\theta$, the price of a European-style option on a ZCB, $P(T,S)$, with option's expiry, $T$, strike, $K$, and maturity of the underlying bond $S$, is given by:
\begin{eqnarray}
\label{eqn:optionZCB_rHW}
V^{\text{Z}}_{\chi}(t,T,S,K;\vartheta)=\sum_{n=1}^N\omega_nV^{\text{Z}}_{\chi}(t,T,S,K;\theta_n),
\end{eqnarray}
where,
\begin{eqnarray}
\nonumber
V^{\text{Z}}_{\chi}(t,T,S,K;\theta_n)&=&\E^\Q_{t}\left[\frac{M(t)}{M(T)}\max(\chi(P(T,S;\theta_n)-K),0)\right]\\
&=&\chi P(t,S)F_{\mathcal{N}(0,1)}(\chi d_n)-\chi KP(t,T)F_{\mathcal{N}(0,1)}(\chi(d_n-\bar\sigma_n)),
\label{eqn:ZCBOption}
\end{eqnarray}
with $P(t,S)$ and $P(t,T)$ the ZCB computed from the associated yield curve; $P(T,S;\theta_n)$ is the ZCB obtained from the model with the randomized parameter, $\vartheta$;  $\chi=1$ and $\chi=-1$ corresponds to call and put options, respectively. Depending on the randomized parameter, we have:
\begin{eqnarray}
&&\text{for}\;\;\eta:\;\bar\sigma^2_n=\frac{\eta_n^2}{2\lambda}\left(1-\e^{-2\lambda(T-t)}\right)B^2(T,S;\lambda),\\
&&\text{for}\;\;\lambda:\;\bar\sigma^2_n=\frac{\eta^2}{2\lambda_n}\left(1-\e^{-2\lambda_n(T-t)}\right)B^2(T,S;\lambda_n),
\end{eqnarray}
with
\[d_n=\frac{1}{\bar{\sigma}_n}\log\frac{P(t,S)}{P(t,T)K}+\frac{\bar\sigma_n}{2},\;\;\;B(T,S;\lambda)=\frac{1}{\lambda}\left(1-\e^{-\lambda(S-T)}\right),\]
where the HW model is defined in~(\ref{eqn:HW_SDE}) and $F_{\mathcal{N}(0,1)}(\cdot)$ corresponds the standard normal CDF.
\begin{proof}
The proof is a direct consequence of combining Theorem~\ref{thm:rHW_generic} with the pricing of options on a ZCB under the HW model (as given in~\cite{BrigoMercurio:2007,OosterleeGrzelakBook}).
\end{proof}
\end{lem}

Now, we consider an interest rate swap with a fixed rate, $K$, and payment times $\mathcal{T} =\{T_i,T_{i+1},\dots, T_m\}$ and the corresponding reset rates $\{T_{i-1},T_{i+1},\dots, T_{m-1}\}$  with the payoff given by:
\[H^{Swap}_{\text{P/R}}(\mathcal{T},K)=\bar\alpha\sum_{k=i}^m\tau_k\left(\ell(T_{k-1};T_{k-1},T_{k})-K\right),\]
for $\tau_k=T_{k}-T_{k-1},$ with $P$ indicating a swap payer for $\bar\alpha=1$ and a swap receiver for $\bar\alpha=-1$ and where $\ell(t;T_{k-1},T_k)$ stands for the libor rate over the period $[T_{k-1},T_k]$ observed at time $t.$

To determine today's value of the swap, we evaluate the corresponding expectation of the \index{discounted cash flow} discounted future cash flows, i.e., each payment which takes place at the time points, $T_{i},\dots,T_m$, needs to be
discounted to today,
\begin{eqnarray}
\label{eqn:swap}
V_{\text{P/R}}^{\text{Swap}}(t,K,\mathcal{T})=\bar\alpha \sum_{k=i}^m\tau_k\E^\Q_t\left[\frac{M(t)}{M(T_k)}\big(\ell_k(T_{k-1})-K\big)\right]=\bar\alpha\sum_{k=i}^m\tau_kP(t,T_k)\big(\ell_k(t)-K\big),
\end{eqnarray}
with $\ell_k(t):=\ell(t;T_{k-1},T_k).$

We derive the valuation formula for a swaption contract with strike $K$ and  option expiry $T=T_{i-1}$. Using~(\ref{eqn:swap}), the pricing equation reads:
\begin{eqnarray*}
	V^{\text{Swpt}}_{\text{P/R}}(t,T,K,\mathcal{T})=\E^{\Q}_t\left[\frac{M(t)}{M(T)}\max\left(V_{\text{P/R}}^{\text{Swap}}(T,K,\mathcal{T}),0\right)\right]=P(t,T)\E^{T}_t\left[\max\left(V_{\text{P/R}}^{\text{Swap}}(T,K,\mathcal{T}),0\right)\right],
\end{eqnarray*}
 where the expectation, $\E^T_t[\cdot]$, is taken under the $T-$forward measure and where the price of a swap price at $T=T_{i-1}$ is given by:
\begin{eqnarray*}
V_{\text{P/R}}^{\text{Swap}}(T,K,\mathcal{T})=\bar\alpha\left[1-P(T,T_m)-K\sum_{k=i}^m\tau_k
P(T_i,T_k)\right]=\bar\alpha-\bar\alpha\sum_{k=i}^mc_kP(T_i,T_k),\label{eqn:c_k_def}
\end{eqnarray*}
 with $c_{k}=K\tau_k$ for $k=i,\dots,m-1$, $c_m=1+K\tau_m$, and $\tau_k=T_k-T_{k-1}$.

So, the pricing equation becomes:
\begin{eqnarray*}
V^{\text{Swpt}}_{\text{P/R}}(t,T,K,\mathcal{T})=\bar\alpha P(t,T)\E^{T}_t\left[\max\left(1-\sum_{k=i}^mc_kP(T_i,T_k),0\right)\right].
\end{eqnarray*}

Up to this point, the pricing equations do not depend on specific model choices. Now, however, we consider pricing under the rHW model, which, via the conditional expectation approach, as presented in Lemma~\ref{thm:rHW_generic}, yields:
\begin{eqnarray}
\label{eqn:swaptionInner}
V^{\text{Swpt}}_{\text{P/R}}(t,T,K,\mathcal{T};\vartheta)&=&\bar\alpha P(t,T)\E_t^T\left[\E^{T}_t\left[\max\Big(1-\sum_{k=i}^mc_kP(T_i,T_k),0\Big)\Big|\vartheta=\theta\right]\right]\\
&=&\bar\alpha P(t,T)\sum_{n=1}^N\omega_n\E^{T}_t\left[\max\Big(1-\sum_{k=i}^mc_kP(T_i,T_k;\theta_n),0\Big)\right]+\epsilon_N.\nonumber
\end{eqnarray}
Here, $\epsilon_N$ is the associated quadrature error defined in~(\ref{eqn:RAnDError}). Since the inner expectation in~(\ref{eqn:swaptionInner}) resembles the expression for a swaption under the classical HW, we follow the standard procedure and apply the so-called Jamshidian trick~\cite{Jamshidian:1989:ExactBond} to exchange the maximum operator and the expectation. The resulting pricing equations for swaptions under the rHW model are given in Lemma~\ref{lem:swaption_rHW}.
\begin{lem}[Pricing of Swaptions under randomized Hull-White model]
\label{lem:swaption_rHW}
Consider the rHW model, with parameters $\{\lambda,\eta\}$ and the randomizing random variable $\vartheta$, which randomizes either of the model parameters. For a unit notional, a constant strike, $K$, option expiry $T=T_{i-1}$ and a strip of swap payments $\mathcal{T} =\{T_i,\dots,T_m\}$, with $T_i>T_{i-1}$ and accruals $\tau_i=T_i-T_{i-1}$, the prices of swaption payer and receiver, $\text{P/R}:=\text{Payer/Receiver}$, are given by:
\begin{eqnarray}
\label{eqn:swaptionPricing_rHW}
V_{\text{P/R}}^{\text{Swpt}}(t_0,T,\mathcal{T},K;\vartheta)=\sum_{n=1}^N\omega_n\sum_{k=i}^mc_kV^{\text{Z}}_{\chi}(t_0,T,T_k,\hat{K}_k(\theta_n);\theta_n),
\end{eqnarray}
with a swaption payer, $\text{P}$, for $\chi=-1$, swaption receiver, $\text{R}$, with $\chi=1$, where $V^{\text{Z}}_{\chi}(\cdot)$ is defined in~(\ref{eqn:ZCBOption}) and where the strike price $\hat{K}_k(\theta_n)=\exp\left(A(T,T_k;\theta_n)+ B(T,T_k;\theta_n)r^*_n\right)$. Here, $r^*_n$ is determined by
solving, for each parameter realization $\theta_n$, the following equation:
\begin{equation}\label{eqn:Jamsh_eqn}
1-\sum_{k=i}^mc_k\exp\Big(A(T,T_k;\theta_n)- B(T,T_k;\theta_n)r^*_n\Big)=0,\;\;\;n=1,\dots,N,
\end{equation}
where
\begin{eqnarray*}
A(T,T_k;\{\lambda,\eta\})&=&\log\frac{P(0,T_k)}{P(0,T)}+ B(T,T_k;\{\lambda,\eta\})f(0,T)-\frac{\eta^2}{4\lambda}\left(1-\e^{-2\lambda T}B^2(T,T_k;\{\lambda,\eta\})\right),\\
B(T,T_k;\{\lambda,\eta\})&=&\frac{1}{\lambda}\left(1-\e^{-\lambda(T_k-T)}\right),
\end{eqnarray*}
with  $c_{k}=K\tau_k$ for $k=i,\dots,m-1$, $c_m=1+K\tau_m$ and where the pairs $\{\omega_n,\theta_n\}$, $n=1,\dots,N$, are based on the randomizer $\vartheta$ and computed using~\ref{res:zeta}.
\end{lem}

Under the rHW model, the pricing Equation~(\ref{eqn:swaptionPricing_rHW}) for swaptions is a direct application of Theorem~\ref{thm:rHW_generic}. The pricing under randomized parameters is simply an {\it average} of non-randomized prices, with varying model parameters, accompanied by weights based on the randomizer.

From the computational perspective, the swaption pricing in~(\ref{eqn:swaptionPricing_rHW}) requires $N$ swaption prices for different realizations $\theta_n$, $n=1,\dots,N$; therefore, $N$ optimization problems in~(\ref{eqn:Jamsh_eqn}) need to be solved. This computational complexity can be greatly reduced by employing the multi-d Newton-Raphson algorithm to determine optimal $r^*_n$ for $n=1,\dots,N.$

\begin{rem}[Computation of sensitivities]
The computation of sensitivities under the RAnD method is straightforward, i.e., because the pricing is expressed as a convex combination, the sensitivities are expressed as a weighted sum of individual derivatives. For example, in the case of the swaption pricing, the sensitivity to a particular market quote $q$ is expressed as a sum of sensitivities of individual options on ZCBs:
\begin{eqnarray}
\frac{\partial}{\partial q}V_{\text{P/R}}^{\text{Swpt}}(t_0,T,\mathcal{T},K;\vartheta)=\sum_{n=1}^N\omega_n\sum_{k=i}^Mc_k\frac{\partial}{\partial q}V^{\text{Z}}_{\chi}(t_0,T,T_k,\hat{K}_k(\theta_n);\theta_n),
\end{eqnarray}
with the specification as given in Lemma~\ref{lem:swaption_rHW}.

On the other hand, the sensitivity to the quadrature pairs, $\{\omega_n,\theta_n\}$, $\partial \theta_n/\partial \hat a$ and $\partial \omega_n/\partial \hat a$ may be, for some specific cases, computed analytically (see Section~\ref{sec:calibration_MarketData}). Still, in a generic setting, it is recommended to compute these derivatives numerically, with, for example, finite differences:
\[\frac{\zeta(\vartheta(\hat a+\delta_{\hat a}))-\zeta(\vartheta(\hat a-\delta_{\hat a}))}{2\delta_{\hat a}}\approx \left\{\frac{\partial \omega_n}{\partial \hat a},\frac{\partial \theta_n}{\partial \hat a}\right\},\]
where $\vartheta(\hat a)$ indicates the dependence of the random variable and parameter $\hat a$, $\delta_{\hat a}$ is the {\it shock} size and $\zeta(\vartheta):\R\rightarrow \{\omega_n,\theta_n\}_{n=1}^N$ is defined in~\ref{res:zeta}. Due to the applied finite difference shocks to $\hat a$, an additional bias will be introduced. We expect, however, this error to be of acceptable magnitude, as is commonly observed in current financial practice.
\end{rem}
\section{Numerical experiments}
\label{sec:numericalExperiments}
In this section, several pricing experiments will be performed. First, we present a detailed analysis of the rHW model in realistic pricing scenarios. In the first experiment, we explore the implied volatility smile evolution in time, comparing the implied volatility surface from the HW and rHW models. As the next step, the study of the parameter randomization on shapes of implied volatilities will be illustrated. Finally, the calibration results with market data will be presented in the conclusive experiment. This section will end with numerical experiments involving bivariate distribution for the model parameters.

\subsection{Evolution of implied volatilities}
\label{sec:evolutionIR}
We analyze the swaption implied volatilities observed in the market and compare them to implied volatilities generated using the HW and rHW models. We focus here on the flexibility in generating realistic implied volatility shapes. Throughout the section, we will consider the {\it shifted} implied volatilities computed by inverting Black's formula, which, for $\bar\alpha\in\{1,-1\}$ for call and put options, respectively, reads:
\begin{eqnarray}
\label{eqn:BS}
V_{\text{B}}(T,K,F_0,\sigma_s,\bar\alpha,s)&=&\bar\alpha\cdot (F_0+s) \cdot F_{\mathcal{N}(0,1)}(\bar\alpha d_1)-\bar\alpha (K+s)F_{\mathcal{N}(0,1)}(\bar \alpha d_2),\\\nonumber
d_1&=&\frac{1}{\sigma\sqrt{T}}\left[\log (F_0+s)/(K+s)+1/2\sigma_s^2T\right],\\\nonumber
d_2&=&d_1-\sigma_s\sqrt{T},
\end{eqnarray}
with shift parameter $s$, $F_0$ being the forward rate, $\sigma_s$ is the corresponding volatility coefficient, $K$ being the strike, and $T$ corresponds to the time to option expiry.

Then, the shifted Black's formula for swaptions is given by:
\begin{eqnarray}
\label{eqn:swaptionBlack}
V^{\text{B,Swpt}}_{P/R}=V_{\text{B}}(T,K,S(t_0),\sigma_s,\bar\alpha,s)\sum_{i=1}^m\tau_iP(t_0,T_i),\;\;\;S(t_0)=\frac{P(t_0,T_{i-1})-P(t_0,T_m)}{\sum_{i=1}^m\tau_iP(t_0,T_i)},
\end{eqnarray}
with swaption payer, $P$, for $\bar\alpha=1$ and swaption receiver, $R$, for $\bar\alpha=-1$ and where $S(t_0)$ is the corresponding swap rate. The shift parameter, $s$, is typically perceived as the {\it lower bound} for the interest rates by the market participants and varies depending on the currency. It is important to note that when inverting the Black's formula in~(\ref{eqn:swaptionBlack}), the corresponding implied volatility $\sigma_s$ is a function of the shift parameter, i.e., for different choices of $s$, the implied volatilities are different. In principle, this is not a problem as long as the implied volatilities from the market and model volatilities are computed with the same shift coefficient.

\begin{figure}[h!]
  \centering
    \includegraphics[width=0.45\textwidth]{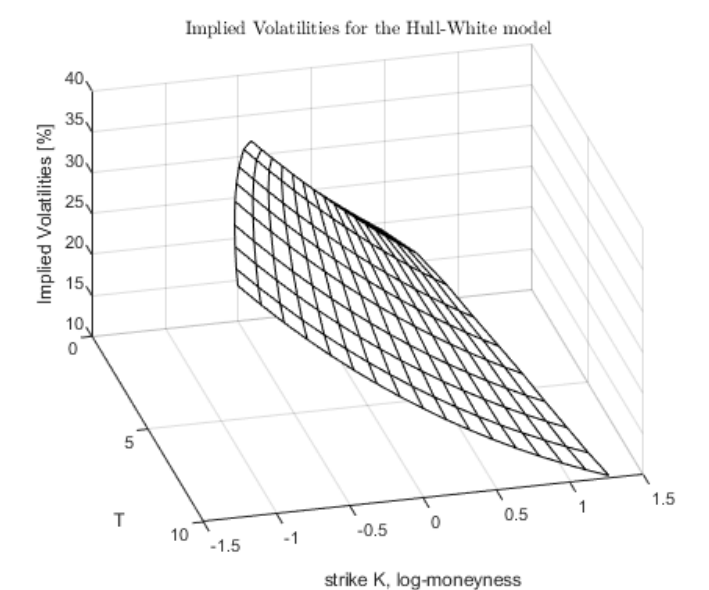}
    \includegraphics[width=0.45\textwidth]{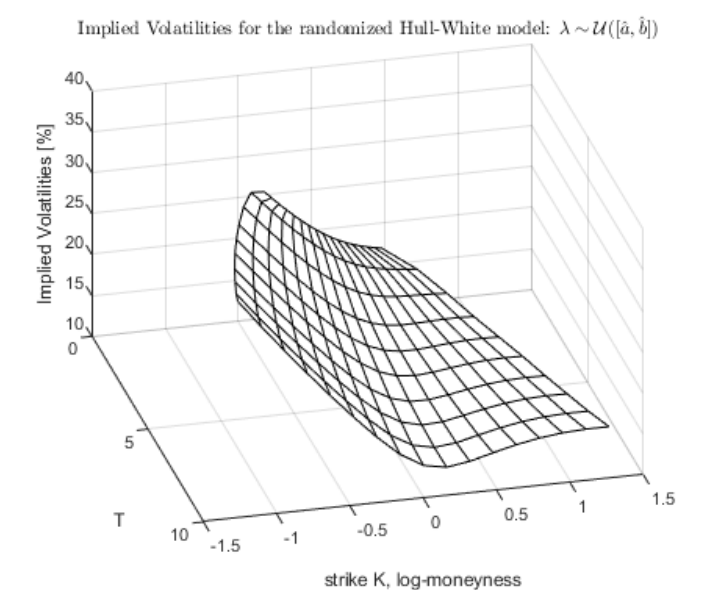}
      \caption{Swaption volatility evolution for the HW and rHW models implied by the shifted Black's model. The simulation was performed for varying swaption option expiry, $T$, and a fixed tenor of $1y$. The parameters specified in the experiment are: for the HW model: $\eta=0.005$, $\lambda=0.001$ and for the rHW model: $\eta=0.005$ and $\lambda~\sim\mathcal{U}([-0.15,0.6]).$ In the experiment, the implied volatilities are computed with zero shift parameter, $s=0.$ }
      \label{fig:3D}
\end{figure}

In the first experiment, we compare the implied volatility surfaces of the HW model and the rHW model. In the rHW model, the speed of mean-reversion is randomized by a uniform distribution, $\lambda\sim\mathcal{U}([\hat{a},\hat{b}])$, on an interval $[\hat{a},\hat{b}].$ The numerical results are illustrated in Figure~\ref{fig:3D}. The results demonstrate that the HW model can only generate implied volatility skew. At the same time, the randomization of the mean-reversion parameter, $\lambda$, shows implied volatility skew and smile. Although the randomization is not {\it time-dependent}, i.e., the parameters are stochastic but stationary, we observe a time evolution of the implied volatilities. The same phenomenon has been observed in~\cite{brigo2002lognormal,grzelakRAnD}, where the randomized Black-Scholes model was considered.
The reported numerical results are promising. By a stochastic mean-reversion parameter, the HW model has an additional degree of freedom that improves the model's flexibility. In the following sections, we will analyze the randomization effect of all model parameters and the different choices of randomizing distributions.

\subsection{Randomization and impact on implied volatilities}
\label{sec:impact_on_IV}
Here, we study the impact of different randomization choices on the implied volatility shape. Understanding how parameters affect the generated volatilities is crucial in model calibration. A clear relationship between the model parameters and the associated skew, curvature and volatility level is fundamental, enabling traders to react swiftly to market movements. It is also the basis for setting up a calibration routine when a particular market movement can be encapsulated in a parameter change.

In the first experiment, we analyze the randomization of volatility parameter $\eta$. We consider two randomization cases where $\eta$ either follows a uniform or a normal distribution. The details regarding the computation of the associated moments and the corresponding quadrature points can be found in~\ref{sec:appendix}. The experiment is set up so that both approaches' mean values, $\E[\eta]$, are equal.
\begin{figure}[h!]
  \centering
    \includegraphics[width=0.45\textwidth]{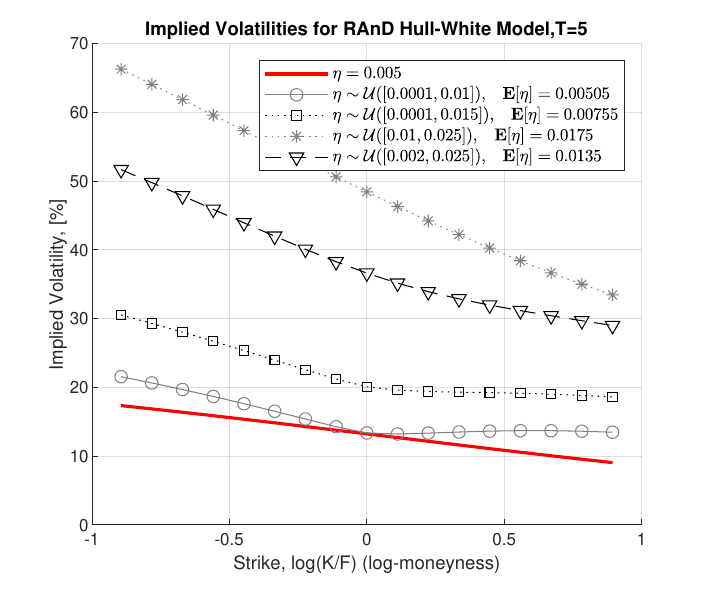}
    \includegraphics[width=0.45\textwidth]{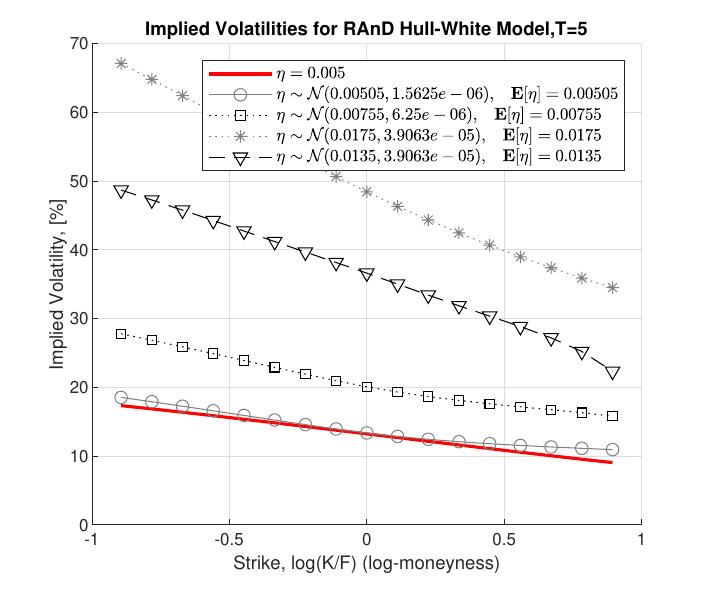}
      \caption{Impact of randomized volatility parameter $\eta$ on implied volatilities with a fixed speed of mean-reversion parameter $\lambda=0.009.$ Left: $\eta\sim\mathcal{U}([\hat{a},\hat{b}])$. Right: $\eta\sim\mathcal{N}(\hat{a},\hat{b}).$}
      \label{fig:impact_1}
\end{figure}
The results are presented in Figure~\ref{fig:impact_1}, where it is shown that the randomization of the volatility parameter $\eta$ has a pronounced effect on the level of implied volatility with minimal effect on the smile. The curvature is visible for uniform randomization. It is essential to note that a higher variance for the normal distribution may give rise to a higher curvature but may also cause issues related to negative volatilities. Our experiments have shown that even for distributions defined in the positive domain, the impact on the smile is limited, even for fat-tailed random variables.

\begin{figure}[h!]
  \centering
    \includegraphics[width=0.45\textwidth]{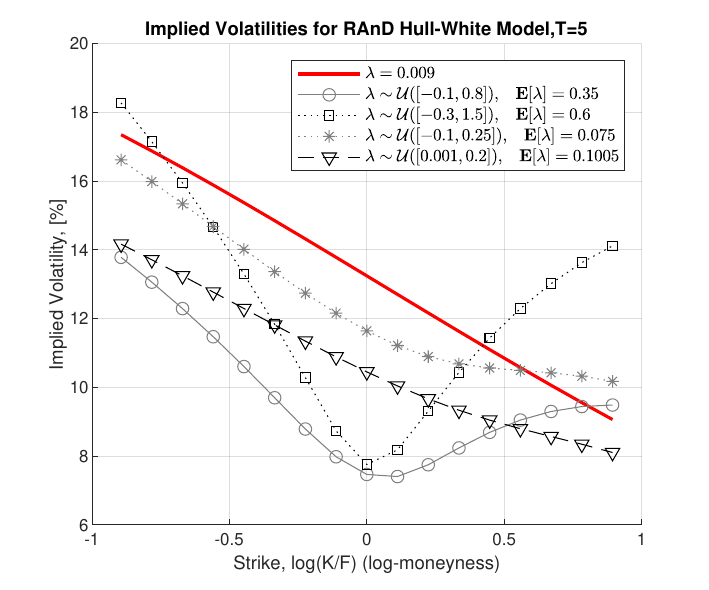}
    \includegraphics[width=0.45\textwidth]{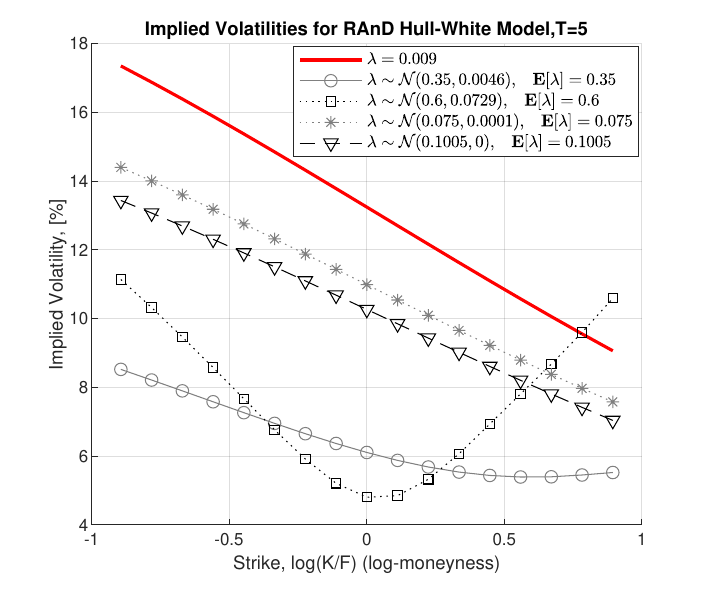}
      \caption{Impact of randomized volatility parameter $\lambda$ on implied volatilities with a fixed volatility parameter $\eta=0.0050.$ Left: $\lambda\sim\mathcal{U}([\hat{a},\hat{b}])$. Right: $\lambda\sim\mathcal{N}(\hat{a},\hat{b}).$}
      \label{fig:impact_2}
\end{figure}

A much richer spectrum of implied volatility shapes is obtained when the randomization technique is applied to the mean-reversion parameter $\lambda$. Figure~\ref{fig:impact_2} presents the randomization with either uniform or normal random variables. A substantial amount of curvature can be generated by taking the mean-reversion random. We also report that the curvature change affects the overall volatility level, i.e., it is impossible to keep the level fixed and only adjust the smile. However, the implied volatility level can be fixed by adjusting the volatility parameter $\eta$. This strategy will be discussed further in the context of model calibration.

\begin{figure}[h!]
  \centering
    \includegraphics[width=0.45\textwidth]{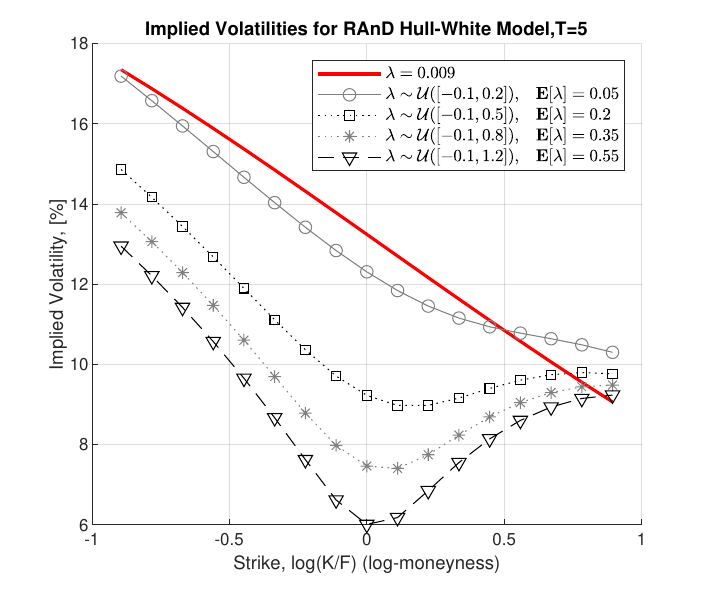}
    \includegraphics[width=0.45\textwidth]{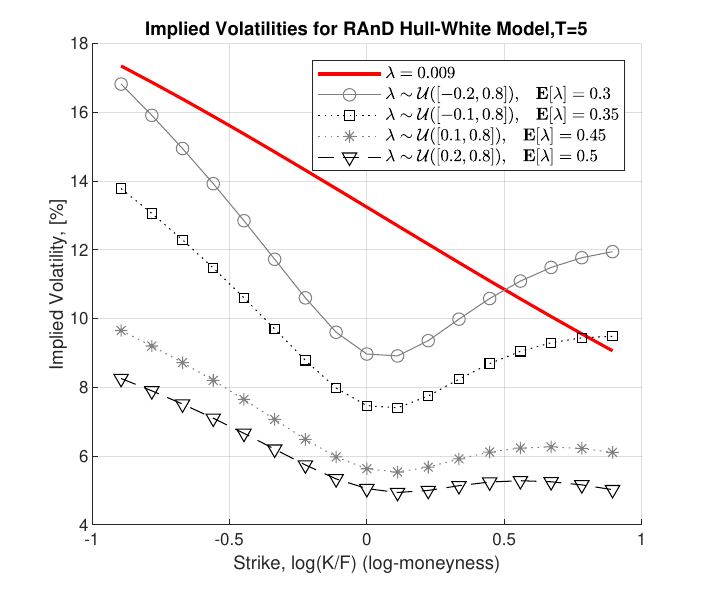}
      \caption{Impact of randomized volatility parameter $\lambda\sim\mathcal{U}([\hat{a},\hat{b}])$ on implied volatilities with a fixed volatility parameter $\eta=0.0050$. Left: varying parameter $\hat{a}$. Right: varying parameter $\hat{b}.$}
      \label{fig:impact_3}
\end{figure}

In the final experiment of this section, we consider the randomization of $\lambda\sim\mathcal{U}([\hat{a},\hat{b}])$ using uniform distribution and check how the parameters $\hat{a}$ and $\hat{b}$ affect the implied volatilities. In Figure~\ref{fig:impact_3}, the results show an interesting pattern: the curvature level is mainly driven by the distance $|\hat{b}-\hat{a}|$, i.e., the larger the distance, the more implied volatility smile is generated. Changes of either of the parameters affects the volatility level; therefore, some of the volatility effect, $\eta$, can be offset by the interval $[\hat{a},\hat{b}].$

Given the numerical results presented above, we conclude that the most considerable impact on implied volatilities comes from the mean-reversion, $\lambda$, and not from the volatility parameter, $\eta.$ This is particularly interesting when we confront these results with Proposition~\ref{prop:rand_eta} and Proposition~\ref{prop:rand_lambda}, where the dynamics of the corresponding short-rate processes were derived and where it was shown that randomization of the mean-reversion did not lead to the local volatility type of dynamics. In contrast, the randomization of $\eta$ does, which can be explained by the nature of interest rate derivatives, where derivative prices are driven by the dynamics of the ZCBs, but not directly by the short-rate process. However, since under the HW model, the volatility of the ZCBs is given by both parameters (see~\cite{OosterleeGrzelakBook}), the randomization of either of them will imply a local volatility type of dynamics for the ZCBs.

\subsection{Calibration of the randomized Hull-White model}
\label{sec:calibration_MarketData}
Every model that may be considered successful must also show its ability to calibrate to the market quotes. Moreover, the calibration and pricing process needs to be computationally efficient, especially when considering the pricing of large portfolios, for example, in the context of xVA. Therefore, this section focuses on calibrating the rHW model to swaption implied volatilities.

Calibration of model parameters requires multiple iterations over the parameter space until specific optimization criteria are met. This implies that the pricing needs to be repeated at every choice of parameter candidate. By randomization of the model parameters, the number of free parameters will increase. The additional degrees of freedom will depend on the type of randomizer chosen. Let us consider, for example, parameter randomization with $\vartheta(\hat a,\hat{b})$, which is driven by two model parameters $\hat a$ and $\hat b$. Two parameters of the randomizer $\vartheta$ correspond to one additional degree of freedom that can be used for model calibration.

From the computational perspective, using the RAnD method requires that for every iteration step in the calibration procedure, the mapping between the randomizing variable $\vartheta$ and the corresponding pairs $\{\omega_n,\theta_n\}$, $n=1,\dots,N$, needs to be employed. Although the algorithm given in~\ref{res:zeta} is straightforward (it mainly depends on the computation of eigenvalues), it can be further simplified. When randomizing variables can be expressed as a linear combination of some {\it base} random variable, for example, normal or uniform, the calculations of the weights and the corresponding nodes can be significantly simplified. In the case of a normal randomizer, $\vartheta\sim\mathcal{N}(\hat a,\hat b^2)$, we can benefit from the linearity of the normal distribution. For the standard normal and its associated points, we have $\theta_n=\hat a + \hat b \cdot \theta_{\mathcal{N},n}$, where $\theta_{\mathcal{N},n}$ are the nodes corresponding to standard normal, $\mathcal{N}(0,1).$ This implies that we can simply {\it tabulate} the results for the standard normal and scale the points accordingly. In the case of the weights $\omega_n$, they stay invariant to a linear transformation~\cite{scmc2019}. A similar property holds for $\vartheta\sim\mathcal{U}([\hat a,\hat b])$, where the nodes can be computed for $\vartheta\sim\mathcal{U}([0,1])$ and scaled appropriately.

In the calibration experiment, we consider the market data for the USD market as of the 18th of August, 2022. In all calibration exercises, we consider a fixed tenor of $1y$ and analyze the accuracy for varying expires and strikes. The calibrated model parameters for both HW and rHW models are tabulated in Table~\ref{Tab:Calibration}. The results are intriguing, i.e., in the randomization for $\lambda$, we were able to calibrate all swaptions while keeping the mean of the randomizer fixed at $0.1$. This shows that having more degrees of freedom is not necessary to improve the calibration results. As in the standard HW model, we have only used two parameters.

\begin{table}[htb!]
\centering\footnotesize
\caption{\footnotesize Calibration of the HW and rHW model: parameters determined in swaption calibration. }
\begin{tabular}{c|c|c||c|c}
\multicolumn{1}{c}{}&\multicolumn{2}{c||}{$\text{Hull-White}$}&\multicolumn{2}{c}{$\text{RAnD Hull-White}$}\\\hline
$T$, expiry &$\eta$&$\lambda$&$\eta$&$\lambda$\\\hline\hline
$1y$&0.0094&0.0090&0.0091&$\lambda\sim\mathcal{N}(0.1,0.45^2)$\\
$2y$&0.0082  &0.0035 & 0.0080&$\lambda\sim\mathcal{N}(0.1,0.33^2)$\\
$5y$&0.0069&0.0020&0.0079&$\lambda\sim\mathcal{N}(0.1,0.16^2)$\\
$8y$&0.0067&0.0095&0.0080&$\lambda\sim\mathcal{N}(0.1,0.12^2)$\\
$10y$&0.0067&0.0090&0.0082&$\lambda\sim\mathcal{N}(0.1,0.11^2)$\\
$15y$&0.0064&0.0080&0.0085&$\lambda\sim\mathcal{N}(0.1,0.09^2)$\\
$20y$&0.0060&0.0080&0.0086&$\lambda\sim\mathcal{N}(0.1,0.08^2)$
\end{tabular}
\label{Tab:Calibration}
\end{table}

\begin{figure}[h!]
  \centering
    \includegraphics[width=0.45\textwidth]{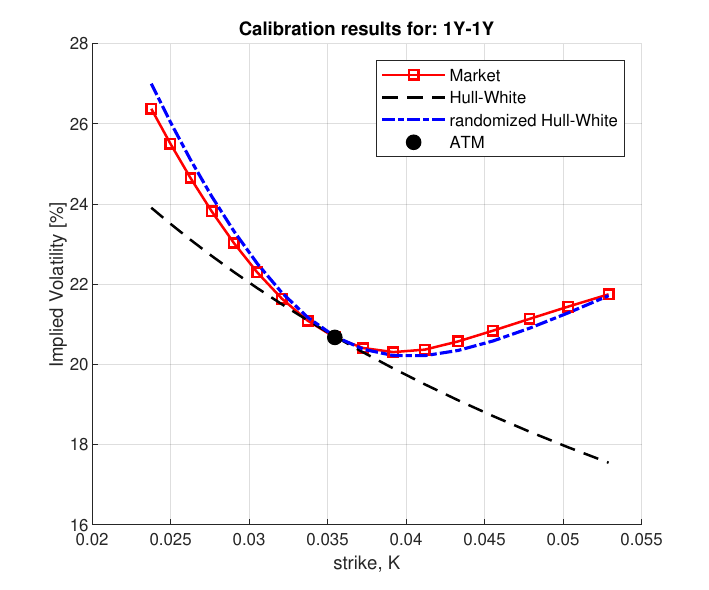}
    \includegraphics[width=0.45\textwidth]{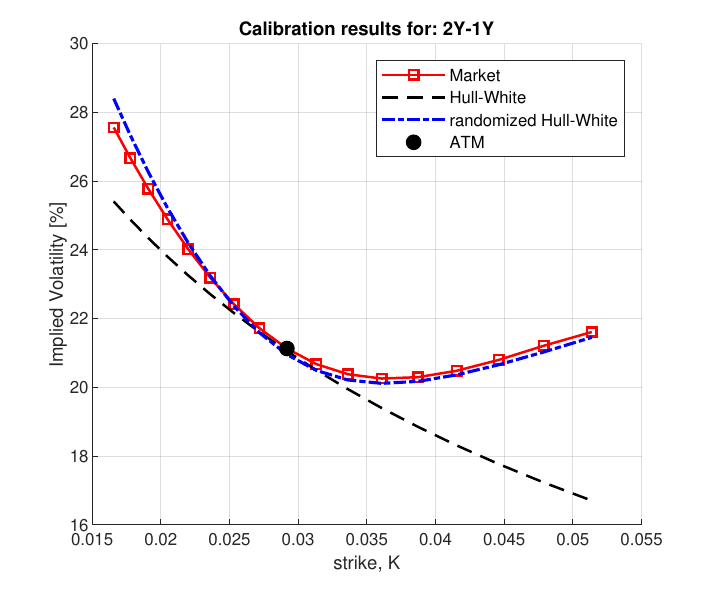}
      \caption{Calibration results of the HW and the rHW models. The market implied volatilities for swaptions were obtained on 18/08/2022 for the USD market. Option expiry: $T=1y$ and $T=2y$ and the implied volatility shift: $s=1\%$. Calibrated parameters are presented in Table~\ref{Tab:Calibration}.}
      \label{fig:calib_1}
\end{figure}

\begin{figure}[h!]
  \centering
    \includegraphics[width=0.45\textwidth]{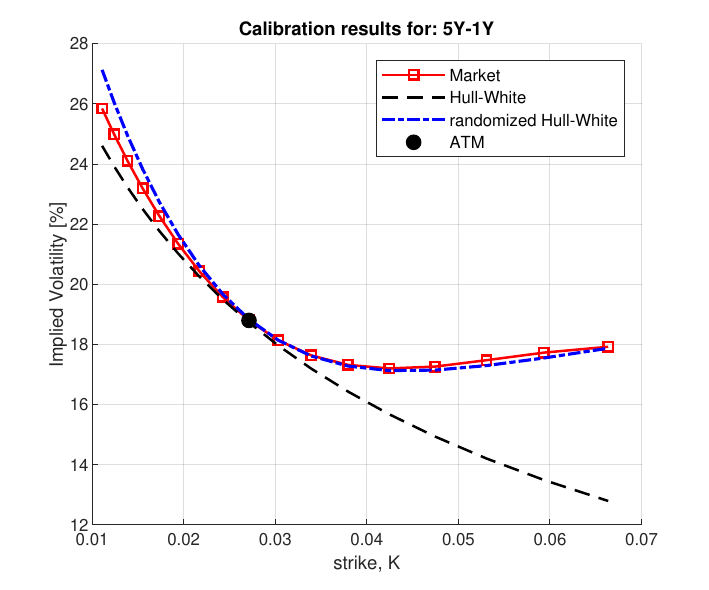}
    \includegraphics[width=0.45\textwidth]{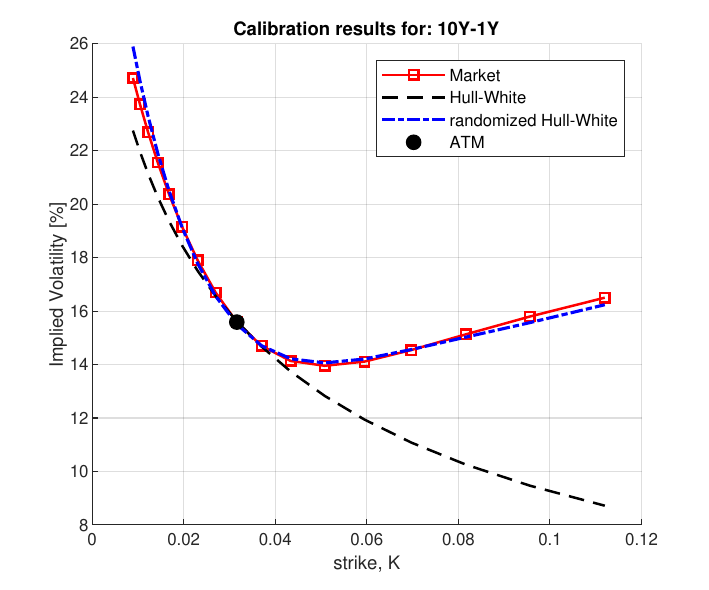}
      \caption{Calibration results of the HW and the rHW models. The market implied volatilities for swaptions were obtained on 18/08/2022 for the USD market. Option expiry: $T=5y$ and $T=10y$ and the implied volatility shift: $s=1\%$. Calibrated parameters are presented in Table~\ref{Tab:Calibration}. }
      \label{fig:calib_2}
\end{figure}

\begin{figure}[h!]
  \centering
    \includegraphics[width=0.45\textwidth]{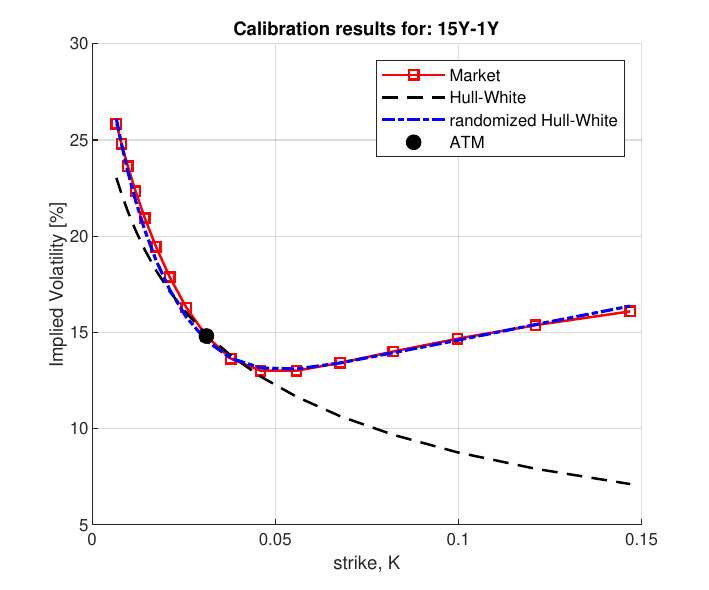}
    \includegraphics[width=0.45\textwidth]{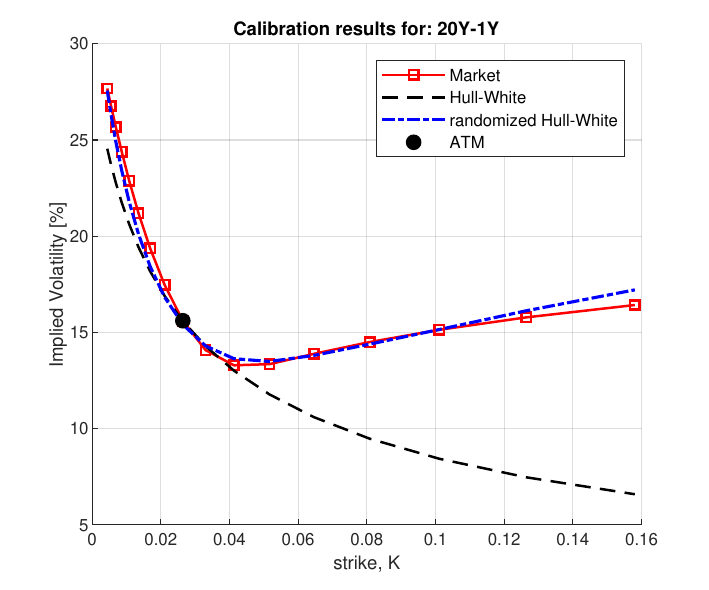}
      \caption{Calibration results of the HW and the rHW models. The market implied volatilities for swaptions were obtained on 18/08/2022 for the USD market. Option expiry: $T=15y$ and $T=20y$ and the implied volatility shift: $s=1\%$. Calibrated parameters are presented in Table~\ref{Tab:Calibration}. }
      \label{fig:calib_3}
\end{figure}

The calibration fit is presented in Figures~\ref{fig:calib_1},~\ref{fig:calib_2} and~\ref{fig:calib_3}. We report an excellent calibration fit for all considered option expiries, varying from 1y to 20y. We have used two parameters in all the calibration cases, just as in the HW model. The results confirm that the RAnD method has great potential for improving existing pricing methods, even with the same number of degrees of freedom.

\subsection{Pricing under bivariate distributions for the model parameters}
\label{sec:pricingBivariate}
This section extends the rHW model and considers a bivariate distribution for both model parameters. Under a bivariate distribution $\Theta=[\vartheta_1,\vartheta_2]$ with $\zeta(\vartheta_1)=\{\omega_{1,n},\theta_{1,n}\}_{n=1}^N$ and conditioned on $\zeta(\vartheta_2|\vartheta_1)=\{\omega_{2,m},\theta_{2,m}\}_{m=1}^M$, the randomized prices are:
\begin{eqnarray}
V_{\rHW}(t,r(t);\vartheta_1,\vartheta_2)=\sum_{n=1}^{N}\omega_{n}\sum_{m=1}^{M}\omega_{m}V_{\HW}(t,r(t);\eta_n,\lambda_m)+\epsilon_{N,M}^b,
\end{eqnarray}
where $N$ and $M$ indicate the number of expansion terms for $\vartheta_1$ and $\vartheta_2|\vartheta_1$, respectively, $\vartheta_2|\vartheta_1$ indicates a conditional random variable, $\epsilon_{N,M}^b$ is the corresponding aggregated error. The remaining specification follows Theorem~\ref{cor:RAnDChF}.

\begin{figure}[h!]
  \centering
    \includegraphics[width=0.45\textwidth]{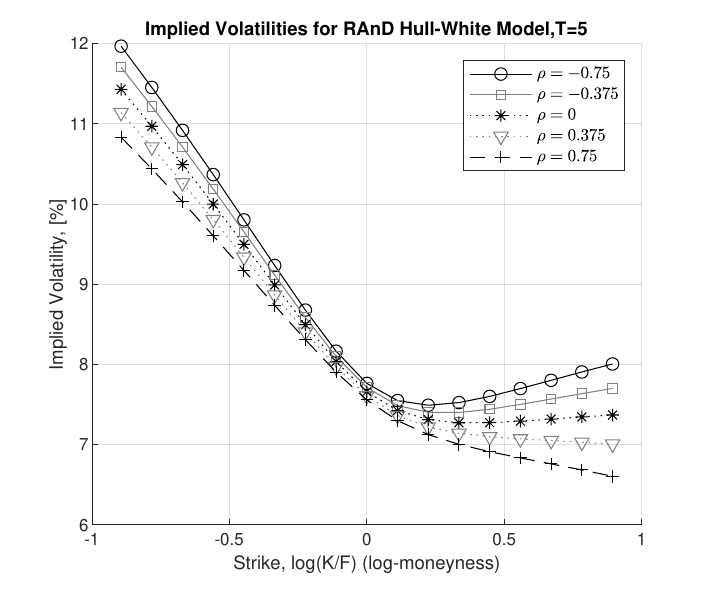}
    \includegraphics[width=0.45\textwidth]{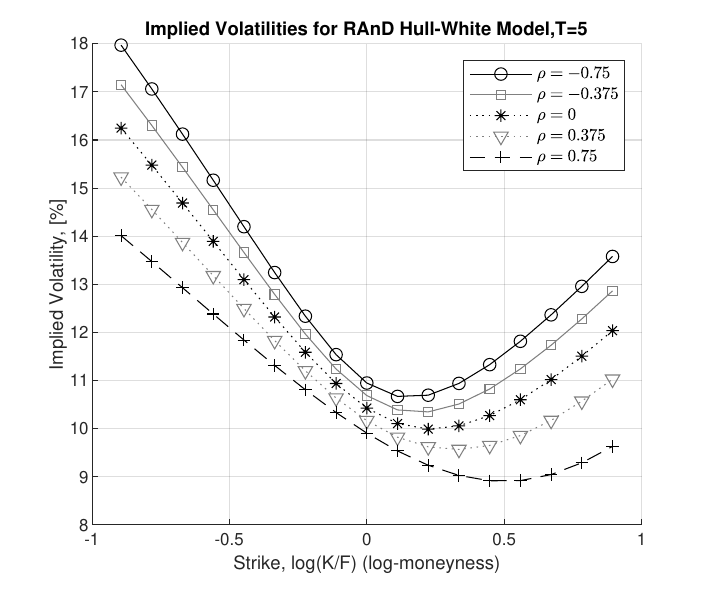}
      \caption{Impact of randomized volatility parameters $\eta$ and $\lambda$ on implied volatilities driven by bivariate normal distribution. Results are presented for varying correlation, $\rho$. Left: $\eta\sim\mathcal{N}(0.008,0.002^2)$ and $\lambda\sim\mathcal{N}(0.5,0.05^2)$. Right: $\eta\sim\mathcal{N}(0.01,0.002^2)$ and $\lambda\sim\mathcal{N}(0.5,0.2^2)$. }
      \label{fig:impact_correlation}
\end{figure}

As an example, let us consider a bivariate normal distribution for the pair $(\vartheta_1,\vartheta_2)$ with the corresponding realizations $(\eta,\lambda)$ for which we have:
\begin{eqnarray}
\vartheta_2|\vartheta_1=\eta_{n}\sim\mathcal{N}\left(\mu_{\lambda}+\frac{\sigma_\lambda}{\sigma_{\eta}}\rho(\eta_n-\mu_{\eta}),(1-\rho^2)\sigma_\lambda^2\right),
\end{eqnarray}
where $\rho$ is the correlation coefficient between $\vartheta_1$ and $\vartheta_2$.
In Figure~\ref{fig:impact_correlation}, we illustrate the impact of the correlation coefficient $\rho$ on the swaption implied volatilities. We report that similar to, e.g., the Heston model, the correlation controls the implied volatility skew. Furthermore, we observe that a higher positive correlation generates more skew, while a negative correlation generates more volatility curvature (smile).

\begin{rem}[Volatility term structure and feasible strategy for model calibration]
Commonly, the volatility-parameter $\eta$ is piece-wise constant, so the ATM volatilities are well calibrated. This strategy will also work with the randomized mean-reversion parameter, i.e., the mean-reversion parameter can be used for smile/skew calibration, while the volatility parameter, $\eta$, will ensure a proper fit of the ATM level.
\end{rem}

\subsection{Convergence results}
\label{sec:convergence}
As presented in Theorem~\ref{thm:rHW_generic}, applying the RAnD method produces a quadrature error. In this section, we analyze the convergence of the error depending on the number of expansion terms, $N$.

\begin{figure}[h!]
  \centering
    \includegraphics[width=0.45\textwidth]{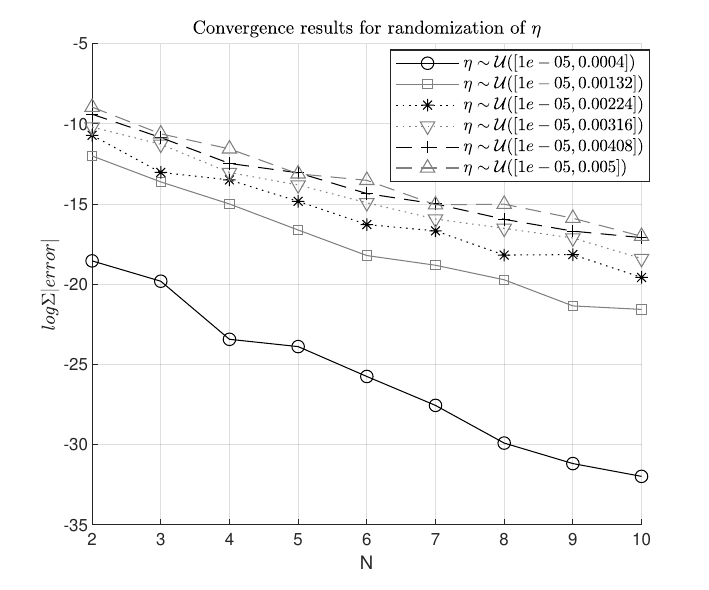}
    \includegraphics[width=0.45\textwidth]{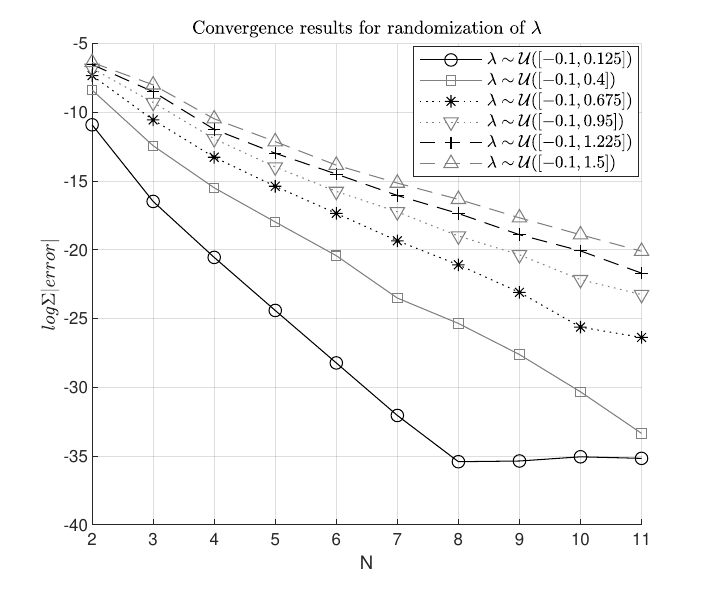}
      \caption{Convergence results for randomization of volatility parameter, and mean reversion parameter $\eta$ and $\lambda$  under the HW model respectively. The base parameters in the experiment were $\eta=0.00625$ and $\lambda= 0.002$.}
      \label{fig:convergence_N}
\end{figure}

The convergence speed depends on the distribution and its parameters, as presented in Figure~\ref{fig:convergence_N}. We report an excellent convergence factor; however, as expected, it depends on the variance of the randomizing random variable. Both cases will achieve satisfactory results for already $N=5$. Randomization using different random variables showed equivalent patterns of convergence.

\section{Conclusion}
\label{sec:conclusion}
In this paper, we have applied the randomization technique to enhance the flexibility of short-rate models for interest rates. We have shown that the model parameters driven by a random variable, instead of being deterministic, facilitate a practical extension of standard, well-popularized models. In addition, this article points out how the normal mixture (a sum of the Hull-White PDFs) can be expressed as a one-dimensional diffusion process with a local volatility function.

We have illustrated that, for the randomized Hull-White model (rHW), one can utilize the available closed-form pricing equations and benefit from flexibility in controlling implied volatilities. In particular, we have shown that the rHW model results in almost perfect swaption calibration.

Finally, the RAnD method is generic and is not limited to any particular modelling choice; therefore, it opens many possibilities for improving existing pricing frameworks.

\bibliographystyle{abbrv}
\bibliography{bibliography.bib}
\appendix
\section{Moments and optimal quadrature pairs, $\{\omega_n,\theta_n\}_{n=1}^N$}
\label{sec:appendix}

\subsection{Moments and optimal quadrature pairs, $\{\omega_n,\theta_n\}_{n=1}^N$}
\label{sec:table}

\begin{table}[htb!]
\centering\footnotesize
\caption{\footnotesize Selected distributions for the stochastic parameters. For the normal random variable with some mean, $\mu$, and variance, $\sigma^2$, it is sufficient to consider standard normal distribution, $\mathcal{N}(0,1)$, and properly scale the $\theta_n$ points, obtained from~\ref{res:zeta}.}
\begin{tabular}{c|c|c}
name&raw moment&domain  \\\hline\hline
$\vartheta\sim\mathcal{U}([\hat a,\hat b])$&$\E[\vartheta^n]=\frac{\hat b^{n+1}-\hat a^{n+1}}{(n+1)(\hat b-\hat a)}$&$[\hat a,\hat b]$ \\
$\vartheta\sim \exp(\hat a)$&$\E[\vartheta^n]=\frac{n!}{\hat a^n}$& $\R^+$\\
$\vartheta\sim\mathcal{N}(0,1)$&$\E[\vartheta^n]=(n-1)!!$\;\text{if}\;\;$n$\;\text{even}; $0$ otherwise&$\R$ \\
$\vartheta\sim\Gamma(\hat a,\hat b)$&$\E[\vartheta^n]=\hat b^n\Gamma(n+\hat a)/\Gamma(\hat b)$&$\R^+$\\
$\vartheta\sim\chi^2(\hat a,\hat b)$&$\E[\vartheta^n]= 2^{n-1}(n-1)!(\hat a+n\hat b)+\sum_{j=1}^{n-1} \frac{(n-1)!2^{j-1}}{(n-j)!}(\hat a+j\hat b )\E[\vartheta^{n-j}]$&$\R^+\cup\{0\}$\\
\end{tabular}
\label{Tab:Moments}
\end{table}

\subsection{Optimal quadrature pairs, $\{\omega_n,\theta_n\}_{n=1}^N$, based on moments}
\label{res:zeta}
Let $\{p_i\}_{i=0}^N$ with $\deg(p_i)=i$ be a sequence of orthogonal polynomials in $L^2$, with respect to PDF, $f_\vartheta(\vartheta)$, of $\vartheta$, then the following holds,
\begin{equation}\label{eqPsi}
\E\left[p_i(\vartheta)p_j(\vartheta)\right] = \int_\R p_i(x) p_j(x)f_\vartheta(x)\d x =
\delta_{i,j}\E\left[p_i^2(\vartheta)\right],\;\;\;i,j=0,\dots,N,
\end{equation}
with $\R$ the support of $\vartheta$, $\delta_{i,j}$ the Kronecker delta.
Orthogonal polynomials $\{p_i\}_{i=0}^N$ satisfy the following recursion relation: \begin{eqnarray*}
{p}_{i+1}(x)&=&(x-\alpha_i){p}_i(x)-\beta_i{p}_{i-1},\;\;\;i=0,\dots,N-1,\\
{p}_{-1}(x)&\equiv& 0,\;\; {p}_0(x)\equiv 1,
\end{eqnarray*}
where $\alpha_i$ and $\beta_i$  are determined in terms of the moments of a random variable $\vartheta$. Consider the monomials $m_i(\vartheta)=\vartheta^i$, and define $\mu_{i,j}$ as $\mu_{i,j}=\E\left[{m}_i(\vartheta){m}_j(\vartheta)\right]$.
From all moments $\mu_{i,j}$, we construct the Gram matrix
$M=\{\mu_{i,j}\}_{i,j=0}^N$, which is symmetric and contains all moments $\{1,\E[\vartheta^1],\dots,\E[\vartheta^{2N}]\}$. Since matrix $M$ is, by
definition, positive definite~\citep{GoWe}, we decompose
$M=R^\T R$, by the Cholesky decomposition of $M.$

The following step relates the Cholesky upper-triangular matrix $R$ to the orthogonal polynomials.  This relationship has been established
in~\citep{GoWe} and is given by,
\begin{eqnarray}
\label{eqn:alphaBeta}
\alpha_j=\frac{r_{j,j+1}}{r_{j,j}}-\frac{r_{j-1,j}}{r_{j-1,j-1}},\;\;\; j=1,\dots,N,\;\;\;\text{and}\;\;\;\beta_j=\left(\frac{r_{j+1,j+1}}{r_{j,j}}\right)^2,\;\;\;j=1,\dots,N-1,
\end{eqnarray}
with $r_{0,0}=1$ and $r_{0,1}=0$ and where $r_{i,j}$ is the $(i,j)$-th element of matrix $R$. Now, we determine the symmetric tridiagonal matrix, $J$,
\begin{eqnarray}
\label{eqn:EigenVal}
{J}:=\left[\begin{array}{ccccc}\alpha_1&\sqrt{\beta_1}&0&0&0\\
\sqrt{\beta_1}&\alpha_2&\sqrt{\beta_2}&0&0\\
 &\ddots&\ddots&\ddots& \\
0&0&\sqrt{\beta_{N-2}}&\alpha_{N-1}&\sqrt{\beta_{N-1}}\\
0&0&0&\sqrt{\beta_{N-1}}&\alpha_{N}\\
\end{array}\right]\in\R^{N\times N},
\end{eqnarray}
with spectral factorization:
\[J=W\Lambda W^T,\;\;\;\Lambda=\text{\rm diag}[\lambda_1^*,\lambda_2^*,\dots\lambda_N^*],\;\;\;WW^\T=I,\]
Due to the positivity of the off-diagonal entries, the eigenvalues $\lambda_j^*$ are distinct, and all $W$'s first row entries are non-vanishing. Moreover, it is well known~\cite{GoWe} that the nodes and weights, $\{x_i,w_i\}_{i=1}^N$, of the Gauss rule are, for $i=1,\dots,N$, given by
$x_i=\lambda_i^*$ and $w_i= ({\bf \epsilon}_1^\T W{\bf\epsilon}_i)^2$,
where ${\bf \epsilon}_j$ is the $i$th axis vector~\footnote{The accompanying Python and MATLAB codes can be found at~\url{https://github.com/LechGrzelak/Randomization}}.


\end{document}